\documentclass[journal]{IEEEtran}
\usepackage[utf8]{inputenc}
\usepackage[LGR, T1]{fontenc} % LGR for greek
\usepackage{textgreek} % for greek
\usepackage{alphabeta} % for greek
\usepackage{microtype}
\usepackage{cite}
\usepackage{hyperref}
\usepackage{graphicx}
\usepackage[caption=false,font=normalsize,
   labelfont=sf,textfont=sf]{subfig}  % Subfigures IEEE style
\usepackage{booktabs}
\usepackage{xcolor}
\usepackage{enumerate}
\usepackage{algorithm}
\usepackage{algorithmic}
\usepackage{orcidlink}

\usepackage{pifont}
\usepackage{bm}
\usepackage{amssymb}
\usepackage{mathtools}
\DeclarePairedDelimiter{\norm}{\lVert}{\rVert}
\DeclareMathOperator*{\argmin}{arg\,min}
\DeclareMathOperator*{\argmax}{arg\,max}
\usepackage{amsthm}
\newtheorem{theorem}{Theorem}
\newtheorem{lemma}{Lemma}
\newcommand*{\transpose}{{\mkern-1.5mu\mathsf{T}}}

\title{A Survey on Archetypal Analysis}
\author{
    Aleix Alcacer$^{\orcidlink{0000-0003-3566-0204}}$,
    Irene Epifanio$^{\orcidlink{0000-0002-6973-311X}}$, 
    Sebastian Mair$^{\orcidlink{0000-0003-2949-8781}}$,
    Morten Mørup$^{\orcidlink{0000-0003-4985-4368}}$
    \thanks{
        Aleix Alcacer and Irene Epifanio are with Jaume I University and valgrAI (I.E.), Spain;
        Sebastian Mair is with Linköping University, Sweden; and
        Morten Mørup is with Technical University of Denmark, Denmark. \\ 
        The author list is alphabetically sorted and all authors contributed equally. \\ 
        Corresponding author: Irene Epifanio, epifanio@uji.es
    }
}
\date{October 2026}

\begin{document}
\maketitle

\begin{abstract}
Archetypal analysis (AA) was originally proposed in 1994 by Adele Cutler and Leo Breiman as a computational procedure for extracting distinct aspects, so-called archetypes, from observations, with each observational record approximated as a mixture (i.e., convex combination) of these archetypes. AA thereby provides straightforward, interpretable, and explainable representations for feature extraction and dimensionality reduction, facilitating the understanding of the structure of high-dimensional data and enabling wide applications across the sciences. However, AA also faces challenges, particularly as the associated optimization problem is nonconvex. This is the first survey that provides researchers and data mining practitioners with an overview of the methodologies and opportunities that AA offers, surveying the many applications of AA across disparate fields of science, as well as best practices for modeling data with AA and its limitations. The survey concludes by explaining crucial future research directions concerning AA.
\looseness=-1
\end{abstract}

\begin{IEEEkeywords}
Archetypal analysis, clustering, convex hull, data science, extreme observations, matrix factorization, survey, unsupervised learning
\end{IEEEkeywords}

\section{Introduction}
\IEEEPARstart{A}{rchetypal analysis (AA)}, originally proposed in \cite{cutler1994archetypal}, is a computational procedure that enables the extraction of the distinct aspects in high-dimensional data as well as how each observation is characterized as a convex combination of these aspects. AA assumes that archetypes (i.e., distinct aspects) can be defined as convex combinations of the collective representation of all data points. AA provides a simple yet interesting computational framework for dimensionality reduction and feature extraction in machine learning with explainable representations that are easily interpretable, as they can be considered idealized pure forms of the data representing archetypical observations.

\begin{figure}[t]
\centering
\includegraphics[width=\columnwidth]{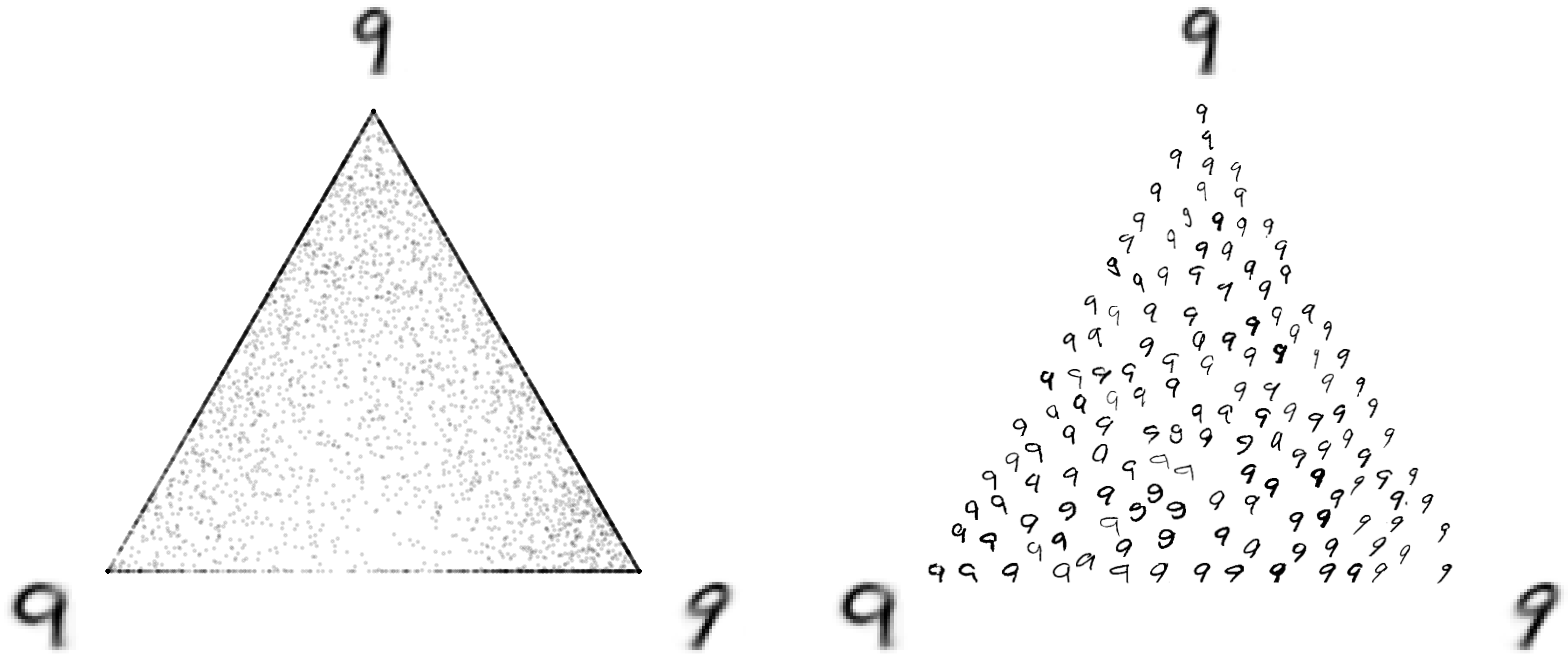}
\caption{AA computed on the subset of the MNIST handwritten digit dataset showing solely the digit~9 with three archetypes. The left part depicts the mixing weights of the nines whereas the right part depicts the actual handwritten digits.
\looseness=-1}
\label{fig:mnist9}
\end{figure}

Figure~\ref{fig:mnist9} shows an example of AA computed on the subset of the MNIST handwritten digit dataset \cite{lecun2010mnist} restricted to the digit~9 with three archetypes. The left part depicts the mixing weights of the 9's, whereas the right part depicts the actual handwritten digits that are represented by those weights. It can be seen that the extracted archetypes of handwritten~9's are \emph{``narrow and straight''}, \emph{``narrow and sloped''}, and \emph{``wide and straight''}. Mathematically, the $28 \times 28$ pixel images depicting digits live in a 784-dimensional space in which we fit three archetypes that span a two-dimensional triangle using AA. The corners of the triangle are the archetypes, and all 9's are projected on this triangle. From the left side, it can be seen that the archetypes themselves, as well as the sides between them, attract more projections than the middle. Furthermore, fewer points are projected on the side between the \emph{``wide and straight''} and \emph{``narrow and sloped''} archetypes.

AA can thereby facilitate the understanding of the structure of high-dimensional data in general providing easily interpretable representations characterizing the data in terms of prominent distinct aspects which can be considered idealized archetypical observations as well as how each observation can be described within a continuum of these aspects by combining these aspects as a convex mixture. However, AA also faces challenges in practical applications. First and foremost, the associated optimization problem is nonconvex, and there is no guarantee of correctly identifying the pure forms, even if they are present as convex combinations of the data. Furthermore, it is unclear how to select the number of archetypes, and in practical applications, outliers can influence the learned representations in undesirable ways. Additionally, the archetypes may be suitable only within a nonlinear manifold, or the data may deviate from the Gaussian normality assumption typically assumed in conventional AA approaches \cite{seth2016probabilistic}.
\looseness=-1

\begin{figure}[t]
\centering
\includegraphics[width=.75\columnwidth]{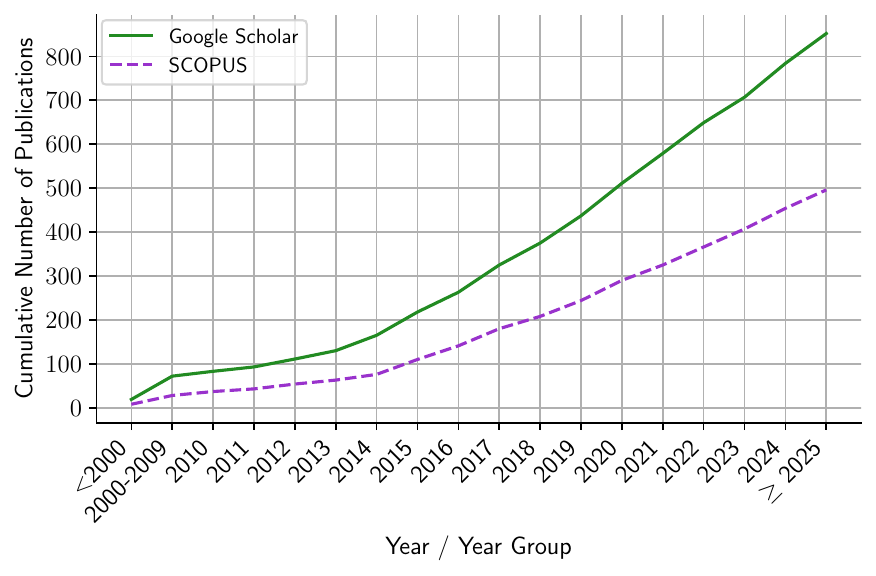}
\caption{Citation overview of paper \cite{cutler1994archetypal} in Google Scholar and SCOPUS until 5th November 2025.}
\label{fig:citations}
\end{figure}

This survey aims to provide researchers and data mining practitioners in general with an overview of the methods and opportunities that AA offers, including an overview of the many disparate fields in which AA has been applied. Furthermore, this survey will highlight best practices for modeling data using AA, identify important limitations of AA, and outline future directions of research advancing AA. Although there was a brief literature review in Portuguese in 2015 developed by \cite{junior2015analise}, this is the first survey of this by now widely adopted methodology for pattern discovery and data modeling; see also Figure~\ref{fig:citations}.
The aim of this survey is also to provide an easily accessible common starting point for this by now large body of works from disparate fields that use and leverage AA to further our understanding of the prominent structures in high-dimensional complex data. The survey thereby also aims to promote accessibility, allowing researchers from other fields to understand and apply AA to new problems by learning from the many problems AA has already addressed. Notably, this survey allows integrating the advances of AA from disparate fields, and systematically documenting and contextualizing them in a single resource. This survey thus also serves as a comprehensive reference, saving researchers the time and effort required to adopt AA as a standard go-to tool.

For the sake of transparency, reproducibility, and thoroughness, and to reduce bias in selecting papers, we searched in the SCOPUS database, the largest online collection of peer-reviewed scholarly works. We selected those works that cite the seminal paper by \cite{cutler1994archetypal} and at the same time have as a keyword \emph{``archetypal analysis''}. A total of 161 documents were found until 5th November 2025. They were incorporated into the different sections of this manuscript or the supplemental file, together with other records identified by their citation in those documents. We further included the most prominent peer-reviewed works obtained from the corresponding Google Scholar search: all the methodological or computational works we found, and those with new applications. 

The survey is structured into the following sections. Section~\ref{sec:History} provides the historical context and concepts behind AA while Section~\ref{Sec:Merits} discusses the merits of AA. Section~\ref{sec:Theory} provides the formal mathematical definition of AA, its theoretical properties, and motivation from different perspectives. Section~\ref{sec:advancemenents2AA} then proceeds by providing the important advancements to AA beyond its original model formulation, whereas Section~\ref{sec:implementations_in_archetypal_analysis} surveys the existing procedures for model inference as implemented in prominent AA software tools. Subsequently, Section~\ref{sec:ProminentApplicationsofAA} provides an overview of the many application domains in which AA has been applied. Finally, Section~\ref{Sec:Limitations} discusses the limitations of AA while Section~\ref{sec:FutureDirandOpenProblems} provides future directions of research and open problems within the field of AA before Section~\ref{sec:Conclusion} concludes this survey.

\section{History of Archetypal Analysis\label{sec:History}}

Archetypes transcend philosophy, psychology, and the sciences in general. The word \emph{archetype} which is derived from the Greek noun ἀρχέτυπον was introduced into the English language in the 1540s with the meaning \emph{``model, first form, original pattern from which copies are made'',} with the adjective archetypal being \emph{``of or pertaining to an archetype''}\footnote{\url{https://www.etymonline.com/search?q=archetype}}. In psychology, Carl Gustav Jung used this concept to define archetypes as \emph{``forms or images of a collective nature''}\footnote{\url{https://www.etymonline.com/word/archetypal\#etymonline_v_26492}}. Archetypes are thus closely related to Plato's pure forms used to define the true nature of things\footnote{\url{https://marcomasi.substack.com/p/platonic-forms-ideas-and-archetypes}}.

In the 1990s, concerns about the interpretability of PCA motivated several constrained factorization models. One notable example, developed independently of AA, is positive matrix factorization (PMF) \cite{paatero1994positive}. PMF enforces nonnegativity constraints on both the factor and weight matrices to obtain interpretable components and predates the later popularization of nonnegative matrix factorization (NMF) \cite{lee1999learning} in the data analysis literature. In contrast to PMF and NMF, AA is rooted in convex geometry and the identification of extreme prototypes.
\looseness=-1

AA emerged from the challenges associated with simulating ozone production in the lower atmosphere, a project supported by the Environmental Protection Agency in the USA. The complex computer models used for this purpose contained hundreds of chemical equations and typically required extensive computation time, often taking 24 hours to simulate just 24 hours of real-time. As a result, researchers could only model a few days of data in each project. This limitation highlighted the need to select data that represented a few \emph{``prototypical''} days, which ultimately inspired the concept of archetypes as a means to effectively capture and analyze key patterns within the dataset. This seminal problem was analyzed in the pioneering paper by \cite{cutler1994archetypal}.
\looseness=-1

AA has since been used in a wide variety of fields, motivated by its ability to characterize prominent properties observed across different domains. As such, in biological systems, AA can be motivated by evolutionary trade-offs and Pareto-optimality that have been found to drive data to reside within a polytope defining trade-offs between optimally defined organisms (akin to pure forms) for various tasks suitable for survival \cite{shoval2012evolutionary}. Within chemistry, AA can be motivated by the observation that measurements of samples from a confined space define concentration fractions in terms of the constituents \cite{morup2012archetypal}, and in geoscience, that hyperspectral images can be represented in terms of distinct spectra (so-called endmembers), such that the recorded spectrum can be defined as convex combinations of these pure spectral forms~\cite{keshava2002spectral}. To extract these pure components, prominent endmember extraction methodologies have focused on minimizing the volume of the data representation \cite{zhuang2019regularization}. As opposed to conventional volume-minimization approaches striving to encapsulate the data manifold from the outside, AA can be considered an approach that encapsulates the data manifold similarly from the inside.
\looseness=-1

Finally, in data science in general, clustering is widely used to identify prototypes, but of interest is also the identification of distinct characteristics in the data and how each observation can be described in terms of these characteristics in easy-to-explain ways \cite{hastie01statisticallearning}. AA also relates to fuzzy clustering approaches in which observations are not hard-assigned but members of multiple groups in a soft manner \cite{mendes2018study,nascimento2019unsupervised,suleman2021comparing}, providing a soft representation of clustering memberships in terms of a continuum.
\looseness=-1

\section{Merits of Archetypal Analysis}\label{Sec:Merits}

Within the family of unsupervised learning techniques, AA provides a unique and complementary perspective on data representation. Unlike principal component analysis (PCA), independent component analysis (ICA), nonnegative matrix factorization (NMF), or $k$-means clustering, AA identifies \emph{extreme} and interpretable prototypes of data, so-called \emph{archetypes}, and represents each observation as a convex combination of these extremes. This geometric formulation offers many advantages.

\textbf{Faithfulness to Observed Data.}
Archetypes are themselves convex combinations of observations, meaning they are constrained to lie in the convex hull, where the convex hull is the minimal convex set enclosing all data points. Thus, all resulting representations stay within the feasible data domain. Hence, this property ensures semantic and physical plausibility and fosters a simplified interpretation and downstream analysis. In contrast, the widely adopted method NMF does not guarantee that latent factors lie within the feasible data domain.

\textbf{Interpretability and Explainability.}
AA provides built-in interpretability: Since the mixing coefficients are convex combinations, they can be easily interpreted probabilistically, directly quantifying similarity to extreme factors, offering a transparent and intuitive description of the data. This property is particularly valuable in application domains that require explainable models, e.g., biomedicine, climate science, and social sciences. Interpretability is even stronger for archetypoid analysis, where the archetypes coincide with actual observations.
\looseness=-1

\textbf{Revealing Extremes and Trade-offs.}
AA is perfectly suited for uncovering boundary structure and trade-offs in data. Archetypes are located, by design, on the boundary of the convex hull of data. Thus, AA highlights Pareto-like relationships and functional extremes, e.g., performance vs. cost or competing biological traits. Such insights support exploratory data analysis, hypothesis generation, and the study of optimality in engineering.

\textbf{Geometric Insight.}
Geometrically, AA can be seen as an approximation of the data's convex hull, helping practitioners to understand global patterns in high-dimensional data. The archetypes then form the vertices of a polytope that encapsulates the majority of observations by minimizing the overall projection error, providing a compact and interpretable summary.

\textbf{Versatility Across Modalities.}
Although vanilla AA is designed for continuous vector data, formulations for kernel~\cite{morup2012archetypal}, probabilistic~\cite{seth2016probabilistic}, functional~\cite{Epifanio2016}, and deep~\cite{keller2021learning} AA exist. Thus, AA can handle a broad spectrum of data types.

\textbf{Complementarity to Clustering and Matrix Factorization.}
AA identifies \emph{extreme} representatives (archetypes), while clustering identifies \emph{typical} representatives (centroids or medoids). Thus, AA provides a complementary view of the data structure. In contrast to NMF or PCA, AA's components are grounded in convex geometry rather than orthogonality or nonnegativity constraints, improving interpretability in many settings.

\textbf{Compact and Interpretable Representations.}
The convex-combination structure facilitates intuitive visualization (e.g., simplex or barycentric plots), making AA especially appealing for communicating results to nonexpert audiences. Its low-dimensional geometric summaries can support data reduction while maintaining transparency.

\section{Theory/Technical description:}\label{sec:Theory}

\begin{figure}[b]
\centering
\includegraphics[width=.9\columnwidth]{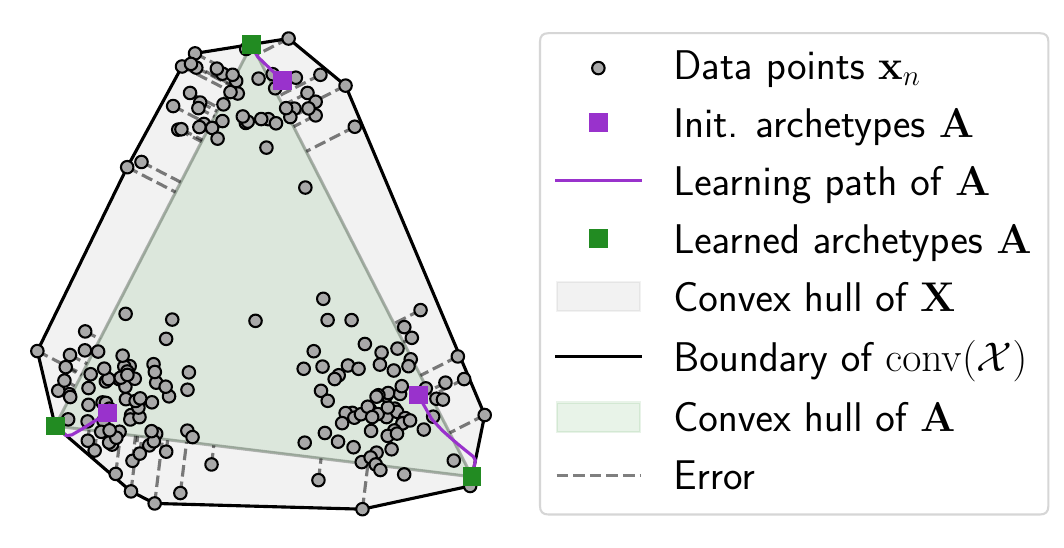}
\caption{An example of AA with $K=3$ archetypes on two-dimensional toy data. After initializing the archetypes (purple squares), the optimization procedure pushes the archetypes (green squares) to lie on the boundary of the convex hull of data (see Theorem~\ref{thm:background:cutler}). The overall objective is to minimize the sum of projection errors (i.e., the sum of all dashed lines).}
\label{fig:AAinit}
\end{figure}

\textbf{Archetypal Analysis (AA).}
Let $\mathcal{X}=\{\mathbf{x}_1, \mathbf{x}_2, \ldots, \mathbf{x}_N\}$ be a dataset consisting of $N$ data points $\mathbf{x}\in\mathbb{R}^M$ in $M$ dimensions and $K$ be the number of archetypes.
Those data points are stacked as row vectors in a design matrix $\mathbf{X} \in \mathbb{R}^{N \times M}$.
The goal is a decomposition of $\mathbf{X}\approx\mathbf{S}\mathbf{A}$ into a product of a weight matrix $\mathbf{S} \in \mathbb{R}^{N \times K}$ and a factor matrix $\mathbf{A} \in \mathbb{R}^{K \times M}$.
The idea of AA is to represent every data point $\mathbf{x}_n$ ($n=1,2,\ldots,N$) as a convex combination of $K$ archetypes $\mathbf{a}_1,\ldots,\mathbf{a}_K$, i.e.,
\begin{align*}
\mathbf{x}_n \approx \mathbf{A}^\transpose \mathbf{s}_n, \quad \text{where}\ \mathbf{s}_n \geq 0 \ \text{and}\ \|\mathbf{s}_n\|_1=1.
\end{align*}
Note that the vectors $\mathbf{s}_n \in \Delta^{K-1} \coloneq \{\mathbf{s}\in\mathbb{R}_{\geq0}^{K} | \sum_{j=1}^K s_j = 1 \}$ contain the weights of the convex combinations which are stacked row-wise in $\mathbf{S}$ and that $\mathbf{A}$ is the design matrix of the archetypes $\mathbf{a}_1,\ldots,\mathbf{a}_K$.
The archetypes are also represented as convex combinations, but this time of data points, i.e.,
\begin{align}\label{eq:factors}
\mathbf{a}_k = \mathbf{X}^\transpose \mathbf{c}_k, \quad \text{where}\ \mathbf{c}_k \geq 0 \ \text{and}\ \|\mathbf{c}_k\|_1=1,
\end{align}
where---just as before---the vectors $\mathbf{c}_k \in \Delta^{N-1}$ are stacked row-wise in $\mathbf{C} \in \mathbb{R}^{K \times N}$ and contain the weights of the convex combinations.
Thus, we have $\mathbf{A}=\mathbf{C}\mathbf{X}$.
Due to the convexity constraints, both weight matrices $\mathbf{S}$ and $\mathbf{C}$ are row-stochastic. For notational convenience, we also index $\mathbf{S}$ column-wise and write
$\mathbf{s}_k\in\mathbb{R}_{\geq0}^N$ for its $k$th column.
A two-dimensional example of AA with $K=3$ is depicted in Figure~\ref{fig:AAinit}.

\textbf{Learning.}
The weight matrices can be found by minimizing the so-called residual sum of squares (RSS) as an objective function, which is given by
\begin{align}\label{eq:AA_LS_objective}
\operatorname{RSS} = \|\mathbf{X}-\mathbf{S}\mathbf{C}\mathbf{X}\|_\text{F}^2 = \|\mathbf{X}-\mathbf{S}\mathbf{A}\|_\text{F}^2.
\end{align}
Here, $\|\cdot\|_\text{F}^2$ denotes the squared Frobenius norm.
Note that optimizing the $\operatorname{RSS}$ to find the optimal weight matrices $\mathbf{S}$ and $\mathbf{C}$ results in a nonconvex optimization problem.
However, the optimization problem is convex in $\mathbf{S}$ for a fixed $\mathbf{C}$ and vice versa.
This yields the standard alternating optimization approach as outlined in Algorithm~\ref{alg:AA}.
Figure~\ref{fig:AAinit} shows the optimization path (purple line) from initialized archetypes (purple squares) to optimized archetypes (green squares).

\begin{algorithm}[t]
    \caption{Alternating optimization of archetypal analysis}
    \label{alg:AA}
    \begin{algorithmic}
        \renewcommand{\algorithmicrequire}{\textbf{Input:}}
        \renewcommand{\algorithmicensure}{\textbf{Output:}}
        \REQUIRE data matrix $\mathbf{X} \in \mathbb{R}^{N \times M}$, number of archetypes $K \in \mathbb{N}$
        \ENSURE factor matrices $\mathbf{S} \in \mathbb{R}^{N \times K}$ and $\mathbf{C} \in \mathbb{R}^{K \times N}$, where $\mathbf{A} = \mathbf{C} \mathbf{X} \in \mathbb{R}^{K \times M}$ and $\mathbf{S} \mathbf{C} \mathbf{X} \approx \mathbf{X}$
        \vspace{1mm}
        \STATE $\mathbf{A} \leftarrow $ initialization of archetypes $\mathbf{A}$ \COMMENT{See Section~\ref{sub:strategies_for_archetypes_initialization}}
        \WHILE{not converged}
            \FOR{$n = 1,2,\ldots,N$}
                \STATE $\mathbf{s}_n = \underset{ \| \mathbf{s}_n \|_1 = 1, \ \mathbf{s}_{n} \geq \bm{0} }{ \argmin } \  \| \mathbf{A}^\transpose \mathbf{s}_n - \mathbf{x}_n \|_2^2$
            \ENDFOR
            \FOR{$k=1,2,\ldots,K$}
                \STATE $\tilde{\mathbf{X}}=\mathbf{X}-\sum_{k^\prime\neq k}\mathbf{s}_{k^\prime}\mathbf{c}_{k^\prime}\mathbf{X}$\\
                \STATE $\mathbf{c}_k = \underset{ \| \mathbf{c}_k \|_1 = 1, \ \mathbf{c}_{k} \geq \bm{0} }{ \argmin } \  \| \tilde{\mathbf{X}}-\mathbf{s}_k\mathbf{c}_k\mathbf{X}  \|_\text{F}^2$
            \ENDFOR
        \ENDWHILE
    \end{algorithmic}
\end{algorithm}

\begin{figure}[t]
\centering
\includegraphics[width=\columnwidth]{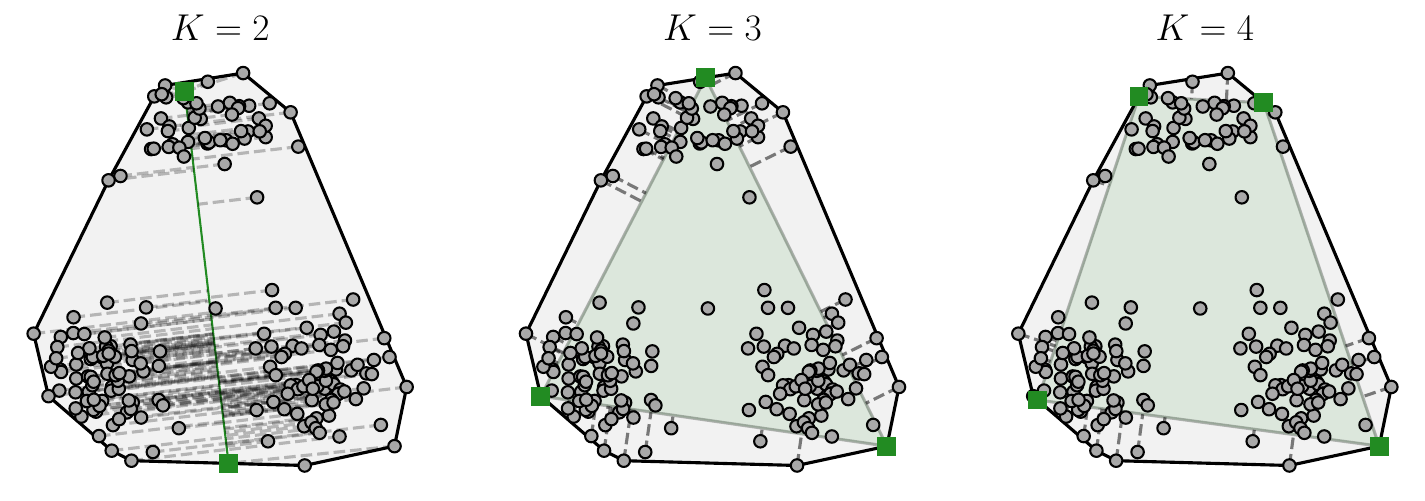}
\caption{An example of AA in two dimensions for various numbers ($K=2,3,4$) of archetypes $\{\mathbf{a}_1,\ldots,\mathbf{a}_K\}$. The archetypes are always located on the boundary of the convex hull of data.}
\label{fig:AA234}
\end{figure}

\begin{figure*}[t]
\centering
\includegraphics[width=.8\linewidth]{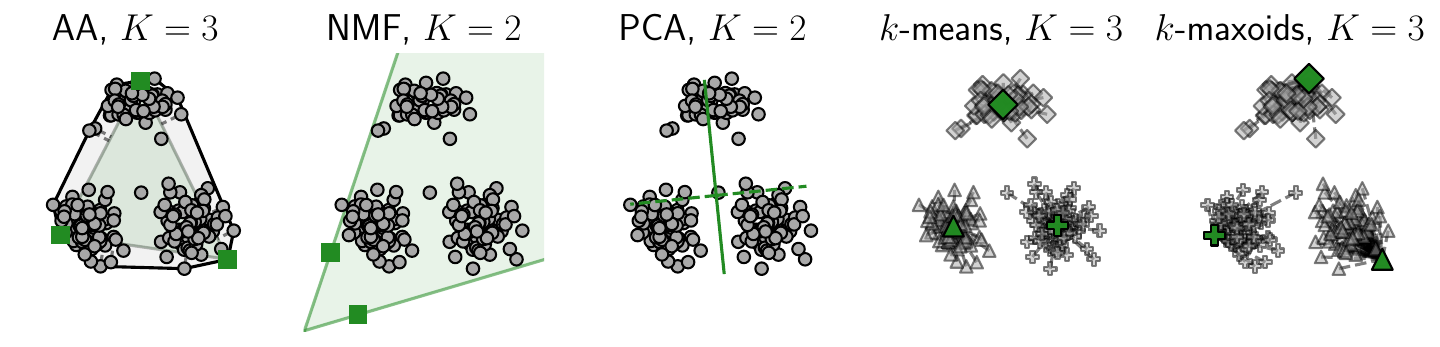}
\caption{AA with $K=3$ archetypes compared to a nonnegative matrix factorization (NMF) with $K=2$ components, a principal component analysis (PCA) with $K=2$ components (first is solid, second dashed), $k$-means clustering with $K=3$ clusters, and $k$-maxoids clustering with $K=3$ clusters.}
\label{fig:AAcomp}
\end{figure*}

\textbf{Properties.}
Several properties of AA can be shown.
The first result characterizes the location of the archetypes: they are always located on the boundary of the convex hull of data for $K>1$, as shown in Figure~\ref{fig:AA234}.

\begin{theorem}[\cite{cutler1994archetypal}]
\label{thm:background:cutler}
Let $\mathcal{X} \subset \mathbb{R}^M$ be a discrete dataset, $\operatorname{conv}(\mathcal{X})$ be its convex hull and $\bm{\mu} \in \mathbb{R}^M$ be the mean of $\mathcal{X}$.
Furthermore, let $K \in \mathbb{N}$ be the number of archetypes and $\partial\operatorname{conv}(\mathcal{X})$ be the boundary of $\operatorname{conv}(\mathcal{X})$ with $|\partial\operatorname{conv}(\mathcal{X})\cap\mathcal{X}|=B$ points on the boundary.
Then, the following holds.
\begin{enumerate}[(i)]
    \item If $K=1$, choosing $\mathbf{a}_1=\bm{\mu}$ minimizes the $\operatorname{RSS}$;
    \item if $1<K<B$, there is a set of archetypes $\{\mathbf{a}_1,\ldots,\mathbf{a}_K\}$ on the boundary of $\operatorname{conv}(\mathcal{X})$ that minimizes the $\operatorname{RSS}$;
    \item if $K=B$, choosing $\{\mathbf{a}_1,\ldots,\mathbf{a}_K\} = \partial\operatorname{conv}(\mathcal{X})\cap\mathcal{X}$ results in a $\operatorname{RSS}$ of zero.
\end{enumerate}
\end{theorem}

The second result states that affine transformations and scaling of the data do not influence the weight matrices when minimizing the $\operatorname{RSS}$.

\begin{lemma}[\cite{morup2012archetypal}]
\label{lem:background:invariance}
The weight matrices $\mathbf{S}$ and $\mathbf{C}$ are invariant to an affine transformation and scaling of the data $\mathcal{X}$.
\end{lemma}
As such, the AA representation is invariant to centering the data by subtracting the mean of each feature, i.e., applying the centering operator $\mathbf{T}=\mathbf{I}-\tfrac{1}{N}\mathbf{11}^\transpose$ and scaling operator $\alpha$ to form the centered and scaled data $\mathbf{\hat{X}}=\alpha\mathbf{TX}$.

Furthermore, the solution of AA does not suffer from rotational ambiguity.

\begin{theorem}[\cite{morup2012archetypal}]
\label{thm:background:uniqueness}
Assume
\begin{align*}
\forall k \exists n\!: \quad &c_{kn}>0 \ \text{and}\ c_{k'n}=0, &&k \neq k', \\
\text{and}\quad 
\forall k \exists n\!: \quad &s_{nk}>0 \ \text{and}\ s_{nk'}=0, &&k \neq k'.
\end{align*}
Then, the solution of AA is unique and does not suffer from rotational ambiguity, i.e., $\mathbf{X}\approx\mathbf{S}\mathbf{C}\mathbf{X}=\mathbf{S}\mathbf{P}^{-1}\mathbf{P}\mathbf{C}\mathbf{X}=\tilde{\mathbf{S}}\tilde{\mathbf{C}}\mathbf{X}$ such that both $\mathbf{S},\mathbf{C}$ and $\tilde{\mathbf{S}},\tilde{\mathbf{C}}$ are equivalent solutions, where $\mathbf{P}$ is a permutation matrix.
\end{theorem}

\begin{table}[b]
\centering
\caption{An overview of related methods and their restrictions.}
\label{tab:related}
\resizebox{\columnwidth}{!}{%
\begin{tabular}{lllll}
\toprule
Algorithm & \multicolumn{2}{c}{Restrictions on $\mathbf{C}$ and $\mathbf{A}$} & \multicolumn{2}{c}{Restrictions on $\mathbf{S}$} \\
\midrule
PCA            & $\mathbf{C}$                             &  
               & $\mathbf{S}$                             & \\
NMF \cite{lee1999learning}            &                                          & $\mathbf{A} \ge 0$
               & $\mathbf{S} \ge 0$                       & \\
Convex NMF \cite{ding2008convex}          & $\mathbf{C} \ge 0$                       & $\norm{\mathbf{c}_k}_1 = 1$
               & $\mathbf{S} \ge 0$                       & \\
Affine Coding \cite{lee1996unsupervised} &                                          & $0 \leq a_{km} \leq 1$ 
               &                                          & $\norm{\mathbf{s}_n}_1=1$ \\
Convex Coding \cite{lee1996unsupervised} &                                          & $0 \leq a_{km} \leq 1$ 
               & $\mathbf{S} \ge 0$     & $\norm{\mathbf{s}_n}_1=1$ \\
Conic Coding \cite{lee1996unsupervised}  &                                          & $0 \leq a_{km} \leq 1$
               & $\mathbf{S} \ge 0$     & \\ 
AA \cite{cutler1994archetypal}            & $\mathbf{C} \ge 0$                       & $\norm{\mathbf{c}_k}_1 = 1$
               & $\mathbf{S} \ge 0$                       & $\norm{\mathbf{s}_n}_1=1$ \\
ADA \cite{vinue2015archetypoids}           & $\mathbf{C} \in \{0, 1\}$                & $\norm{\mathbf{c}_k}_1 = 1$
               & $\mathbf{S} \ge 0$                       & $\norm{\mathbf{s}_n}_1=1$ \\
Fuzzy clustering \cite{mendes2018study}& $\mathbf{c}_{k} = \mathbf{s}_{k}/\norm{\mathbf{s}_k}_1$ &
               & $\mathbf{S} \ge 0$                       & $\norm{\mathbf{s}_n}_1=1$ \\
$k$-means      & $\mathbf{C}\ge 0$                        & $\norm{\mathbf{c}_k}_1=1$
               & $\mathbf{S} \in \{0, 1\}$                & $\norm{\mathbf{s}_n}_1=1$ \\
$k$-medoids    & $\mathbf{C}\in \{0, 1\}$                 & $\norm{\mathbf{c}_k}_1=1$
               & $\mathbf{S} \in \{0, 1\}$                & $\norm{\mathbf{s}_n}_1=1$ \\
\bottomrule
\end{tabular}
}
\end{table}

\textbf{Geometric perspective.}
The minimization of the $\operatorname{RSS}$ can also be seen from a geometric perspective, which is also shown in Figure~\ref{fig:AAinit}.
The archetypes are chosen to span a large convex hull (green shaded area) such that the per-point projections (grey lines) are minimal.
This can be phrased mathematically as
\looseness=-1
\begin{align}\label{eq:geometric_objective}
\operatorname{RSS}
= \sum_{n=1}^N \min_{\mathbf{x}' \in \operatorname{conv}(\{\mathbf{a}_1,\ldots,\mathbf{a}_K\})} \| \mathbf{x}_n - \mathbf{x}' \|_2^2.
\end{align}
Thus, AA can be seen as finding an approximation (green convex hull) of the convex hull of the dataset (grey convex hull) with a given number of vertices ($K$).

\textbf{Volumetric perspective.}
The geometric perspective suggests that AA is about finding a polytope with $K$ vertices within the convex hull of the data. To reduce the total projection error, this polytope has to \emph{maximize its volume}.
This is also done by a related approach called simplex volume maximization (SiVM)~\cite{thurau2010yes}.
The idea is to sequentially choose points as archetypes such that the volume of the convex hull they induce is maximized.
Unlike standard AA, the chosen archetypes are real data points rather than mixtures thereof, akin to the archetypoid analysis (ADA) \cite{vinue2015archetypoids} procedure described later.
\looseness=-1

Other related methods for matrix factorization can be seen from a \emph{volume minimization} perspective, for example nonnegative matrix factorization (NMF) \cite{lee1999learning}.
Generally, most matrix factorization methods can be seen as finding a decomposition $\mathbf{X} \approx \mathbf{S} \mathbf{A}$ of the design matrix $\mathbf{X}$ into a product of a weight matrix $\mathbf{S}$ and a factor matrix $\mathbf{A}$.
Those methods typically differ in their assumptions on the data, the constraints they impose on the factorization (cf. Table~\ref{tab:related}), and the loss function that is being used.
NMF assumes that the design matrix is nonnegative and constrains the weight and factor matrices to be nonnegative as well.
An example comparing AA and NMF is depicted in Figure~\ref{fig:AAcomp}.
We can see that the factors of NMF open up a cone in which the data can be represented and that the angle between the two factors could be wider, pushing the factors further away from data and thus reducing their interpretability.
Therefore, NMF is often regularized to minimize the volume of the cone that is spanned by the factors \cite{lin2015identifiability,fu2016robust}.
Another approach for the interpretability problem is to constrain the factors $\mathbf{A}$ to be convex combinations of data, i.e., $\mathbf{A}=\mathbf{C}\mathbf{X}$ as in Equation~\eqref{eq:factors}, yielding a convex NMF (CNMF) \cite{ding2008convex}.
The notion of convexity is also used in convex coding \cite{lee1996unsupervised}. However, there, the convexity is imposed only on the mixing weights.
\looseness=-1

\textbf{Clustering perspective.}
Various matrix factorization methods such as NMF can also be used for clustering.
Likewise, AA can be seen as a clustering technique where the archetypes $\mathbf{a}_k$ denote cluster representatives and convex combinations (weights) $\mathbf{s}_n \in \Delta^{K-1}$ associated with the data points can be seen as probabilities of cluster memberships.
Unlike $k$-means clustering \cite{lloyd82}, this yields a soft assignment.
However, $k$-means clustering is indeed related to AA in many ways.
Consider for example its objective function
\begin{align*}
\sum_{n=1}^N \min_{\mathbf{x}' \in \{\mathbf{a}_1,\ldots,\mathbf{a}_K\}} \| \mathbf{x}_n - \mathbf{x}' \|_2^2
\end{align*}
and compare it to the geometric perspective of AA as stated in Equation~\eqref{eq:geometric_objective}.
The difference is the location of the projection: to the convex hull of factors in AA versus to the closest cluster center in $k$-means.
Thus, the objective function of $k$-means upper bounds the objective function of AA \cite{mair2019coresets} which was also exploited for initialization purposes \cite{mair2024archetypal}.
Note that $k$-means can also be phrased as a matrix factorization problem $\mathbf{X} \approx \mathbf{S} \mathbf{C} \mathbf{X}$, where $\mathbf{C}$ denotes convex combinations to model the cluster centers and $\mathbf{S}$ contains a standard basis vector per data point modeling the clustering assignment.
As in AA, the cluster centers/factors are mixtures of data points rather than data points themselves, which hinders their interpretability.
For $k$-means, this is fixed by $k$-medoids whereas for AA, this is fixed by archetypoid analysis (ADA)~\cite{vinue2015archetypoids}.
However, unlike AA, $k$-means places the cluster centers in the interior of the dataset.
This is circumvented in $k$-maxoids clustering \cite{bauckhage2015k}, where the cluster representatives are placed on the boundary of the dataset similar to the FurthestSum initialization procedure proposed for AA \cite{morup2012archetypal}.
Figure~\ref{fig:AAcomp} shows how AA compares to $k$-means and $k$-maxoids clustering.
\looseness=-1

\begin{figure}[b]
\centering
\includegraphics[width=\columnwidth]{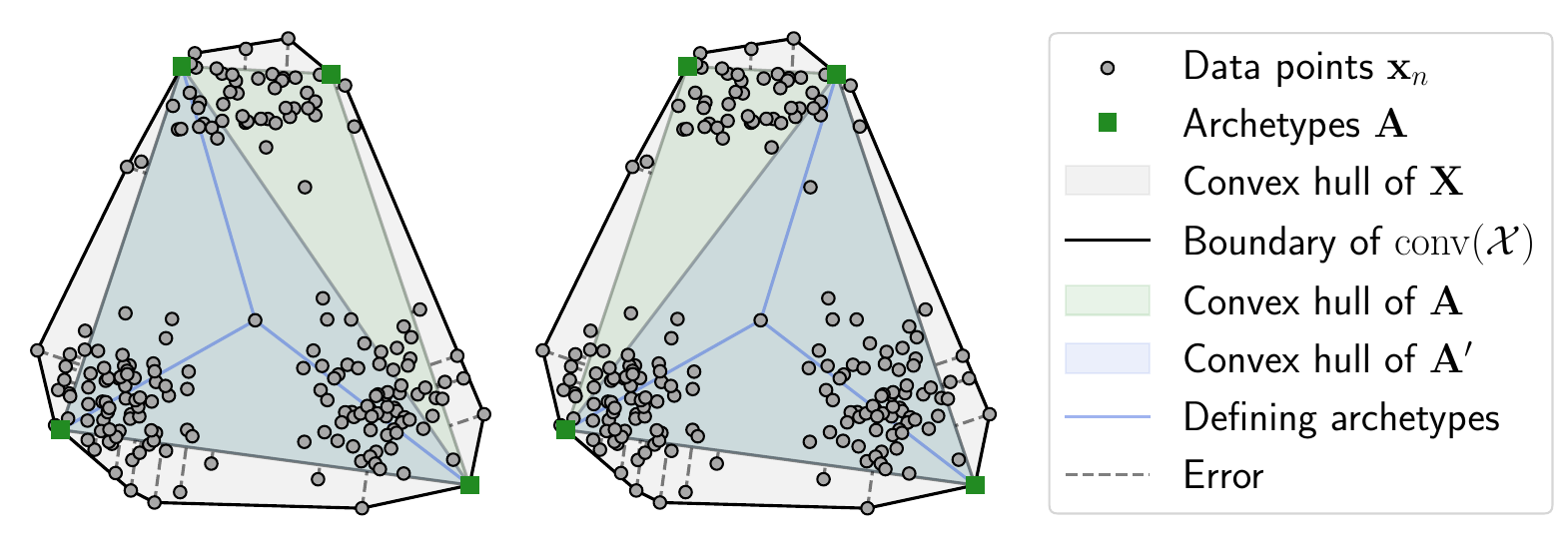}
\caption{Only $M+1$ archetypes $\mathbf{A}'$ are needed to define a data point. Hence, in the case of $K > M+1$, the weights $\mathbf{s}_n$ are not consistent. For the highlighted data point in the middle of the dataset, two different subsets $\mathbf{A}'$ (blue shaded area) of $M+1$ archetypes can be used to define the point. Thus, the representation $\mathbf{s}_n$ is not consistent. Here, $M=2$ and $K=4$.}
\label{fig:AArep}
\end{figure}

\textbf{Change-of-basis perspective.}
AA can also be seen as a change of representation or a change of basis of the data points.
Instead of using linear combinations of the standard basis of a real vector space, AA uses convex combinations of archetypes, thus, representing data $\mathbf{x}_n \in \mathbb{R}^M$ as $\mathbf{s}_n \in \Delta^{K-1}$ on the $(K-1)$-simplex, e.g., as in the left-hand side of Figure~\ref{fig:mnist9}.
This shows another relation of AA to approaches that perform matrix factorization and a change of basis such as independent component analysis~(ICA)~\cite{comon1994independent} and principal component analysis~(PCA).

For pedagogical examples such as in this survey, AA is often depicted for the case of $K > M+1$, where four archetypes ($K=4$) are shown in a two-dimensional ($M=2$) space.
This is atypical in real-world scenarios where we usually deal with high-dimensional data and a relatively low number of archetypes.
The case of $K>M+1$ is problematic since only $M+1$ archetypes are needed to define a point using convex combinations and which archetypes are used often depends on the specific choice of optimizer.
Figure~\ref{fig:AArep} illustrates this problem.
For AA with four archetypes in a two-dimensional space, the point in the middle can be defined using two different subsets $\mathbf{A}' \subset \mathbf{A}$ of size three.
Hence, if representational consistency in the case of $K>M+1$ is needed, the convex combinations $\mathbf{s}_n$ have to be regularized \cite{boubekki2020analysis}.
\looseness=-1

\section{Advancements of archetypal analysis}\label{sec:advancemenents2AA}

AA has been adapted to achieve different goals or to analyze data of different types. Table~\ref{tab:AA_taxonomy} provides a compact comparison of representative extensions according to the nature of the archetypes, their location, the representation space, and their main purpose.

\begin{table*}[t]
\centering
\caption{Taxonomy of representative extensions of archetypal analysis.}
\label{tab:AA_taxonomy}
\scriptsize
\renewcommand{\arraystretch}{0.9}
\setlength{\tabcolsep}{3pt}
\begin{tabular}{p{1.8cm} p{3.2cm} p{3.0cm} p{2.8cm} p{5.4cm}}
\toprule
\textbf{Method} & \textbf{Archetypes} & \textbf{Location} & 
\textbf{Representation space} & \textbf{Purpose / when to use} \\
\midrule

AA 
& Mixtures of observations 
& Boundary of the convex hull 
& Original data space 
& Standard choice for interpretable extremal profiles \\

Relaxed AA 
& Relaxed mixtures of observations 
& May extend outside the convex hull 
& Original data space 
& When extreme profiles may not be well represented by the observed convex hull \\

ADA 
& Actual observations 
& Boundary or interior of the convex hull 
& Original data space 
& When archetypes should correspond to tangible, observed cases \\

Kernel AA 
& Mixtures of mapped observations 
& Boundary of the convex hull in feature space 
& Nonlinear kernel space 
& When nonlinear structure or pairwise similarities are important \\

AAnet / DeepAA 
& Latent archetypal representations 
& Latent archetypal space 
& Learned nonlinear space 
& When the nonlinear representation should be learned jointly with the archetypal structure \\

Sparse AA 
& Sparse mixtures 
& May extend outside the convex hull
& Original data space 
& For sparse representations in hyperspectral unmixing \\

Weighted AA 
& Mixtures of observations 
& Boundary of the convex hull 
& Original data space 
& When observations have unequal importance or weights\\

Robust AA 
& Mixtures of observations 
& Boundary of the convex hull 
& Original data space 
& When robustness to outliers is important \\

Probabilistic AA 
& Archetypal parameter profiles 
& Parameter space 
& Distribution-specific 
& For discrete or non-Gaussian data \\

biAA 
& Mixtures of observations and variables 
& Nested convex hulls 
& Observation--variable space 
& When archetypal structure is sought for both observations and variables \\

\bottomrule
\end{tabular}
\end{table*}

\textbf{Type and location of archetypes.}
An advantage of AA lies in the requirement that the archetypes must be convex combinations of the actual data, ensuring that the representation remains close to the observed data. However, \emph{true} archetypes may not always be accurately represented as convex combinations of the observed data. In this context, \cite{morup2012archetypal} proposed a relaxation of the AA framework to allow for the possibility that archetypes may exist outside the convex hull of the data by changing the original constraint in Equation~\eqref{eq:factors} to $1 - \delta \leq \|\mathbf{c}_k\|_1 \leq 1 + \delta$ for some small $\delta > 0$. 

In some cases, it is essential for the archetypes to be actual observations, meaning they are cases from the sample $\mathcal{X}$. A new archetypal concept, the archetypoid analysis (ADA), was presented to address this issue by \cite{vinue2015archetypoids}. Archetypoids are real data points in the dataset that best represent pure types. In this case, the original constraint in Equation~\eqref{eq:factors} is modified to $\|\mathbf{c}_k\|_1 = 1$, with $c_{kn} \in \{0,1\}$. As a result, the original continuous optimization problem transforms into a mixed-integer optimization problem. In the same vein, \cite{damle2017geometric} worked on separable AA.

\textbf{Nonlinear representation spaces via kernels.}
When the data are better represented in a nonlinearly transformed feature space, a nonlinear AA formulation can be considered.
A major distinction is whether the nonlinear representation is fixed in advance (via a kernel or AA on predefined features) or learned jointly with the archetypes (next subsection).
A first notable work in the area of nonlinear representation spaces is \cite{morup2012archetypal}, which extended AA to data characterized by pairwise relationships, leading to the definition of kernel AA. Notably, kernel AA explores how the AA least squares objective function in Equation~\eqref{eq:AA_LS_objective} can trivially be rewritten using the trace operator $\operatorname{tr}(\cdot)$ so that the data enter only through the Gram matrix
$\mathbf{K}_{\text{lin}}=\mathbf{XX}^\transpose$, i.e.,
\begin{align*}
\|\mathbf{X}-\mathbf{SCX}\|_\text{F}^2
&=\operatorname{tr}(\mathbf{XX}^\transpose-2\mathbf{SC}\mathbf{XX}^\transpose+\mathbf{SC}\mathbf{XX}^\transpose\mathbf{C}^\transpose\mathbf{S}^\transpose)\\
&=\operatorname{tr}(\mathbf{K}_{\text{lin}})-2\operatorname{tr}(\mathbf{SC}\mathbf{K}_{\text{lin}})+\operatorname{tr}(\mathbf{SC}\mathbf{K}_{\text{lin}}\mathbf{C}^\transpose\mathbf{S}^\transpose).
\end{align*}
Since the objective depends on the data only through $\mathbf{K}_{\text{lin}}$, the derivation is agnostic to the choice of Gram matrix: replacing $\mathbf{K}_{\text{lin}}$ with any positive semi-definite kernel matrix $\mathbf{K}$ with entries $K_{ij}=k(\mathbf{x}_i, \mathbf{x}_j)=\langle\phi(\mathbf{x}_i), \phi(\mathbf{x}_j)\rangle_\mathcal{H}$ yields AA in the induced reproducing kernel Hilbert space $\mathcal{H}$, where the archetypes $\mathbf{C}\Phi(\mathbf{X})$ are convex combinations of the mapped observations and the principal convex hull is taken in $\mathcal{H}$ rather than in the input space. Here, $\Phi(\mathbf{X})$ stacks the mapped observations $\phi(\mathbf{x})^\transpose$ row-wise. The gradients with respect to $\mathbf{S}$ and $\mathbf{C}$ likewise depend on the data only through $\mathbf{K}$, so that the alternating optimization schemes from Section~\ref{sec:implementations_in_archetypal_analysis} carry over with $\mathbf{XX}^\transpose$ replaced by $\mathbf{K}$ \cite{zhao2016hyperspectral}.
Whereas kernel AA operates implicitly through inner products, when only dissimilarities are available, \cite{vinue2015archetypoids} instead embed the objects via multidimensional scaling and compute archetypoids in the resulting space.
Kernel AA offers multiple benefits. First, it applies to data available only as pairwise similarities for which no vectorial representation exists (e.g., graphs or sequences). Second, invariances can be encoded in the choice of $k(\cdot,\cdot)$ rather than in the model, and once $\mathbf{K}$ has been computed the cost of the subsequent optimization is independent of the ambient dimension $M$, which is attractive in the high-dimensional $M$ and moderate $N$ regime.
However, the gains come at a cost. An archetype $\sum_i c_{ki}\phi(\mathbf{x}_i)$ has in general no pre-image in the input space. Hence, it cannot be inspected, plotted, or described in terms of the original variables. The remedy is to restrict $\mathbf{C}$ to selection matrices so that archetypes coincide with observed data points, which is precisely the kernel archetypoid formulation of \cite{SUN2017147} and links to archetypoid analysis.
Another problem is that storing $\mathbf{K}$ requires $\mathcal{O}(N^2)$ memory. A remedy is using a Nyström approximation \cite{zhao2016hyperspectral}.

\textbf{Nonlinear representation spaces via neural networks.}
A first family of methods applies AA to representations provided by a fixed nonlinear map stemming from a neural network. \cite{wynen2018unsupervised} apply AA to features from a hidden layer of an image classification network to define artistic styles. The transformation of the space is given a priori, so the archetypes remain well defined relative to a representation that is not itself optimized. Moreover, in explainable AI research \cite{wedenborg2026explaining}, AA is used to interpret the latent feature spaces of neural networks for describing how information is represented layer-by-layer inside the model. 
A second family learns the transformation jointly with the archetypes. \cite{van2019finding} reformulate the objective as $\norm{f(\mathbf{X}) - \mathbf{SA}}_\text{F}^2$, optimizing over both the archetypes $\mathbf{A}$ and an approximately invertible function $f$ realized by an autoencoder (AAnet).
Here, AA is performed on $f(\mathbf{X})$ rather than on $\mathbf{X}$, where $f$ is a learned, nonlinear transformation, typically realized by a neural network.
The DeepAA proposed in  \cite{keller2019deep,keller2021learning} is instead grounded in probabilistic generative models (variational autoencoder model (VAE), deep variational information bottleneck) and \cite{keller2021learning} additionally incorporates side information.
However, a terminological caveat is here needed as the label \emph{deep archetypal analysis} is not exclusively attached to deep-learning-based formulations. A separate line of work applies AA recursively to its own coefficient matrix, obtaining $\mathbf{X} \approx \mathbf{S}_L\mathbf{A}_L\cdots\mathbf{A}_1$ with the $L$ layers fitted greedily \cite{thakur2019deep} and likewise refers to this as deep AA. We refer to it as \emph{stacked AA} to distinguish it from neural-network-based approaches.
To account for both the nonlinearity and the count structure present in scRNA-seq data, \cite{wang2022non} developed scAAnet and \cite{milite2025midaa} proposed MIDAA.

\cite{tasissa2023k} learn a dictionary of landmarks in the data space under a locality-weighted $\ell_1$ penalty, so that each observation is a convex combination of nearby landmarks. The resulting alternating minimization is unrolled into an interpretable autoencoder. Most recently, \cite{wieser2025revisiting} project inputs onto a continuous, low-dimensional hyperspherical manifold, relaxing the simplex assumption underlying AA.
A third, more recent line keeps the nonlinearity in the encoder but re-anchors the archetypes in the observation space. Archetypal sparse autoencoders \cite{fel2025archetypal} retain the archetypal constraint on the archetypes but not on the coefficients, which are produced by a learned sparse encoder. Since optimizing $\mathbf{C}$ is infeasible at their scale ($N>10^8$), the data are first distilled by $k$-means to a candidate set $\widetilde{\mathbf{X}} \in \mathbb{R}^{R \times M}$, $R \ll N$, and the archetypes are formed as
\begin{align*}
  \mathbf{A} = \mathbf{C}\widetilde{\mathbf{X}} + \boldsymbol{\Lambda},
  \qquad \mathbf{c}_k \in \Delta^{R-1}, \quad
  \norm{\boldsymbol{\lambda}_k}_2 \le \delta,
\end{align*}
where $\delta = 0$ enforces membership in $\operatorname{conv}(\widetilde{\mathbf{X}})$ and larger $\delta$ interpolates towards unconstrained dictionary learning. The constraint bounds where archetypes may lie but does not push them outward, and the candidates are cluster centroids rather than extreme points, so the extremal semantics of AA are not inherited. Their stated motivation is stability, since unconstrained sparse autoencoders yield dictionaries that vary substantially across random seeds.

Learning the representation and the archetypes simultaneously compounds an identifiability question that linear AA already has. In the linear case the ambiguity is bounded by the data: the archetypes must lie in $\operatorname{conv}(\mathbf{X})$, so that the fitted polytope is contained in the convex hull of data, though the solution is not unique in general. If $K$ exceeds the number of extreme points, the surplus archetypes may be placed anywhere inside the hull. Similarly, if $f$ is free, the latent geometry carries no absolute frame for reference, and a sufficiently flexible encoder can map essentially any observations onto the vertices of the latent simplex; which points become archetypes is then a property of the learned transformation as much as of the data. \cite{keller2021learning} report that retaining the same molecules but exchanging the side information yields archetypes differing substantially in composition and structure. Existing methods address this in three ways. First, the latent archetype positions are fixed by construction: DeepAA places $K$ archetypes at the vertices of a $(K-1)$-simplex which limits expressiveness. Second, the objective may be regularized stochastically, as in AAnet, where noise added to the latent representation discourages degenerate solutions without providing guarantees. Third, and most consequentially, the model may be supplied with side information: \cite{keller2021learning} argue that typicality is not a property of the data alone but a function of the side information provided, which makes such models semi-supervised in substance even when presented as extensions of an unsupervised method.
\looseness=-1

\textbf{Sparse representations.}
In specific real-world applications, such as hyperspectral unmixing, sparse solutions are essential. This necessity has led to several proposals in this area. \cite{xu2022l} proposed an AA algorithm with $L_1$ sparsity constraint, which also includes the relaxation of AA to archetypes outside the convex hull as proposed in \cite{morup2012archetypal}. Relaxation is also employed by \cite{xu2023manifold} together with an $L_{2,1}$ regularizer to make the unmixed results sparser and a manifold regularizer promoting similarity between neighboring observations within a superpixel representation. Another approach presented by \cite{rasti2023sunaa} consists of transforming the AA formulation into a semi-supervised setting, where the original objective function is changed to $\norm{\mathbf{X} - \mathbf{SCD}}_\text{F}^2$, where $\mathbf{D}$ contains the endmembers provided by a spectral library. 
\looseness=-1

\textbf{Generalizations of archetypal analysis.}
AA only finds archetypes of observations. In \cite{palumbo2012archetypal} AA was used on the variable space instead in order to obtain archetypal variables.
A recent extension combining archetypes of observations with archetypes in the variables is biarchetype analysis (biAA), developed by \cite{alcacerieee24}. BiAA simultaneously identifies archetypes for both observations and features, expressed as mixtures of the biarchetypes. BiAA is to fuzzy biclustering as AA is to fuzzy clustering; and biAA is to AA as fuzzy biclustering is to fuzzy clustering. In biAA, the original optimization problem is replaced by $\norm{\mathbf{X} - \mathbf{SCXRD}}_\text{F}^2$, where the biarchetypes $\mathbf{A}\in \mathbb{R}^{K \times L}$ are given by $\mathbf{A}=\mathbf{CXR}$, i.e., $K$ archetypes for rows and $L$ for columns, with $\sum_{n=1}^N c_{kn} = 1$ and $c_{kn}\geq 0$ and $\sum_{m=1}^M r_{ml} = 1$ and $r_{ml}\geq 0$. Consequently, biarchetypes are built as mixtures of observations and features weighted by $\mathbf{C}$ and $\mathbf{R}$, respectively.
The matrix $\mathbf{D}\in \mathbb{R} ^{L \times M}$ provides the extent to which each biarchetype contributes to the approximation of each feature. Biarchetype analysis has been shown to offer considerable advantages over biclustering methods, especially in terms of interpretability \cite{alcacerieee24}.
\looseness=-1

\textbf{Missing values.}
Like most statistical methods, archetypal analysis assumes that the data is complete. However, incomplete data (characterized by missing values) are frequently encountered in real-world applications. The first proposal that accommodated
missing information was put forward by \cite{morup2012archetypal}, who modified the original objective function. In \cite{EpiIbSi17}, varying weights are assigned to non-missing and missing values to address this issue. In both cases, archetypes could be located outside the convex hull. For that reason, \cite{doi:10.1080/00031305.2018.1545700} proposed several alternatives. In one of them, a dissimilarity matrix based on all pairwise dissimilarities among the data points is calculated and projected to embed the cases within a Euclidean space where AA is conducted. In another alternative called AAII, internal imputations are performed during the parameter updates of AA. Following that line, the approach by \cite{giordani2024weighted} relies on the standard AA method, enhanced by an additional step that imputes missing entries grounded in the weighted least squares method.
\looseness=-1

\textbf{Other data forms.}
AA was originally defined for continuous data; however, recent developments have enabled its application to a wider range of data. \cite{seth2016probabilistic} proposed a probabilistic extension of the AA framework (PAA). The idea is to construct the convex hull within the parameter space and obtain archetypal profiles within it. \cite{seth2016probabilistic} specifically addressed the cases of Bernoulli, Poisson, and multinomial probability distributions. When the observation model follows a multivariate normal distribution with an identity covariance, this formulation is equivalent to solving the original AA problem. \cite{EugsterPAMI} expanded that framework to encompass the general case of nominal observations. Recently, \cite{wedenborg2025archetypal} explored how second-order Taylor expansions of likelihood functions admit the use of state-of-the-art least squares solvers for PAA as exemplified using the Bernoulli distribution for binary data. \cite{kaufmann2015copula} considered another approach and employed a copula-based method to ensure that AA remains unaffected by strictly monotone transformations of the input data. The rationale behind this is that such transformations should generally not impact the identification of points as archetypes. This approach allows the observations to be continuous and/or discrete and have missing values. An alternative method for binary data was proposed by \cite{sort20}, who identified archetypoids in the observation space. As a result, the archetypal patterns derived from ADA represent feasible solutions. This approach eliminates the need for posterior binarization with PAA and enables a more accurate recovery of archetypal information. Similarly, \cite{Italia25} proposed using ADA for nominal observations. As regards ordinal data, \cite{FERNANDEZ2021281} developed a two-step method for applying ADA to ordinal responses based on the ordered stereotype model, while \cite{wedenborg2024modeling} proposed a direct optimization framework for AA for ordinal data. Furthermore, \cite[Chapter~8]{tesisAleix} expanded AA for high-dimensional data.
\looseness=-1

AA has also been extended to other data objects. Specifically, AA has been defined for symbolic data \cite{DEsposito2013}. In particular, \cite{corsaro2010archetypal} and \cite{Esposito2012} have generalized archetypes to interval-valued data, and \cite{santelliarchetypal} to histogram-valued data. Furthermore, AA and ADA have also been defined for dense functional and sparse functional data in \cite{Epifanio2016} and \cite{VinEpi17}, respectively. Previously, \cite{cutler1997moving} presented a variant of AA designed to monitor dynamic structures, such as traveling waves or solitons, and \cite{stone1996archetypal} applied AA to dynamical systems. Additionally, archetypal networks were determined by \cite{Rago15} when networks are the data objects, while AA for relational data defined by signed weighted networks was proposed by \cite{nakis2023characterizing}. Moreover, archetypal shapes with landmarks were studied by \cite{EpiIbSi17}, whereas \cite{epifanio2023archetypal} focused on archetypal shapes of open curves. Furthermore, \cite{olsen2022combining} considered directional AA for axially symmetric data.

\textbf{Weighted archetypal analysis.}
In the original AA problem, each observation---and consequently each residual---contributes equally to the solution. However, in the weighted archetypal problem defined by \cite{Eugster2010}, the original optimization function is replaced by $\norm{\mathbf{W}(\mathbf{X} - \mathbf{SCX})}_\text{F}^2$, where $\mathbf{W}$ is an $N \times N$ matrix of weights. Weighted AA allows for the incorporation of additional information from the dataset, such as the importance of observations or the correlation between them. Moreover, a variant of weighted AA is also needed when using coresets. Coresets are weighted subsets of the complete dataset on which models perform provably competitively compared to operations on all
data, enabling to learn the same AA model faster and on less data \cite{mair2019coresets}. 

\textbf{Robustness.}
\cite{Eugster2010} proposed a way to robustify AA by weighting the residuals and observations respectively, considering a kind of M-estimator for multivariate real-valued data (M-variate), where the domain of their loss function is $\mathbb{R}^M$. In robust analysis, the domain of the loss function is usually $\mathbb{R}^+$, and this is used by \cite{chen:hal-00995911} and \cite{SUN2017147} with the Huber family of loss functions. In \cite{Moliner2019} the Tukey biweight or bisquare family of loss functions that can better cope with extreme outliers was further used for robust multivariate and functional AA.

\textbf{Other variants of AA.}
Various other variants of AA have been developed for solving specific problems. For instance, hierarchical convex-hull nonnegative matrix factorization, which was designed for clustering, was proposed by \cite{kersting2010hierarchical}. Additionally, \cite{ragozini2017archetypal} introduced a data-driven approach for identifying prototypes that integrates AA with compositional data analysis. Furthermore, \cite{hinrich2016archetypal} extended AA to accommodate the modeling of multiset data considering multi-subject fMRI data, which is referred to as multi-subject AA (MS-AA). That model enforces a shared archetype-generating matrix across subjects while permitting subject-specific archetypes and mixing matrices. 

On a theoretical level, diverse contributions have addressed different issues. \cite{doi:10.1137/20M1331792} established the consistency and convergence of AA under bounded support assumptions. \cite{craig2024wasserstein} considered an alternative formulation of AA based on the Wasserstein metric and obtained the existence of solutions in different situations. \cite{javadi2020nonnegative} proposed a formulation of nonnegative matrix factorization balancing the reconstruction error with the distance between the archetypes and the data's convex hull, wherein a special case of this formulation corresponds to AA. The computational and robustness properties of this formulation are further studied in \cite{behdin2024sparse}. Another formulation that has AA as a special case and was developed by \cite{de2019near} is near-convex AA. Here, a parameter regulates the maximum distance between the basis vectors and the convex hull of the data points. 
\looseness=-1

\section{Implementations of archetypal analysis}\label{sec:implementations_in_archetypal_analysis}

There are many inferential approaches to estimating the AA model. Below we highlight the most prominent and provide a summary of their computational complexity in Table~\ref{tab:algorithms}, followed by practical guidance on how to choose among them.

\textbf{Quadratic Programming (QP) using Nonnegative Least Squares (NNLS).}
In their seminal paper \cite{cutler1994archetypal}, the AA model was solved in its least squares formulation, considering alternately solving for $\mathbf{S}$ and $\mathbf{C}$ as described in Algorithm \ref{alg:AA}. Notably, each alternating subproblem defines two quadratic programming (QP) problems 
\begin{align*}
\tfrac{1}{2}\mathbf{z^\transpose Hz}+\mathbf{gz}\quad \text{s.t.}\quad \mathbf{Iz}\geq\mathbf{0},\quad \mathbf{1^\transpose z}=1,
\end{align*}
respectively, given by
$\mathbf{H}=\mathbf{CX}(\mathbf{CX})^\transpose$ and $\mathbf{g}=-\mathbf{x}_n^\transpose(\mathbf{CX})^\transpose$ when solving for $\mathbf{s}_n$ whereas $\mathbf{H}= (\mathbf{s}_k^\transpose \mathbf{s}_k)(\mathbf{X}\mathbf{X}^\transpose)$ and $\mathbf{g}=-\mathbf{s}_k^\transpose\tilde{\mathbf{X}}\mathbf{X}^\transpose$ (where $\tilde{\mathbf{X}}=\mathbf{X}-\sum_{k^\prime\neq k}\mathbf{s}_{k^\prime}\mathbf{c}_{k^\prime}\mathbf{X}$) when solving for $\mathbf{c}_k$. Elegantly, the linear constraint $\mathbf{1^\transpose z}=1$ can be directly imposed using a simple quadratic regularization penalty $\tfrac{\lambda}{2}(\mathbf{1^\transpose z}-1)^2$ where the strength $\lambda$ of the regularization is set very large \cite{cutler1994archetypal,bell2007modeling}. This removes the linear constraint, whereas the regularization can be absorbed into the QP objective by redefining $\mathbf{H}_\lambda=\mathbf{H}+\lambda\mathbf{1}\mathbf{1}^\transpose$ and $\mathbf{g}_\lambda=\mathbf{g}-\lambda\mathbf{1}^\transpose$. Consequently, when solving for $\mathbf{s}_n$, this is achieved by minimizing
\begin{align*}\tfrac{1}{2}\|\mathbf{A}^\transpose\mathbf{s}_n-\mathbf{x}_n\|_2^2+\tfrac{\lambda}{2}\|\mathbf{1}^\transpose\mathbf{s}_{n}-1\|_2^2,
\end{align*}
in which the elements of $\mathbf{s}_n$ are nonnegative and $\lambda$ is set very large to ensure the regularization towards the simplex is satisfied. Notably, this can be solved directly using the nonnegative least squares (NNLS) active set procedure described in \cite{lawson1995solving,bro1997fast}; for details see also \cite{wedenborg2025archetypal}. Solving for the above QP scales cubically in the size of the Hessian matrix $\mathbf{H}$ specified over the active set. When solving for $\mathbf{S}$ assuming the active set uses all $K$ archetypes this becomes for the $\mathbf{S}$-update $\mathcal{O}_{\mathbf{S}}(NK^3)$ whereas when solving for each of the $K$ columns in $\mathbf{C}$ one at a time this scales in the size of the active set $|\mathcal{A}|$ (that is typically much smaller than $N$) as $\mathcal{O}_{\mathbf{C}}(K|\mathcal{A}|^3)$ for the $\mathbf{C}$ update. Instead of relying on NNLS, other convex optimization procedures and software can be used that directly accommodate linear and nonnegativity constraints, such as CVXOPT\footnote{\url{https://cvxopt.org/}} as considered in \cite{pymf}.
\looseness=-1

\begin{table}[t]
    \caption{Comparison of the principal optimization algorithms used in AA. Costs are per update and exclude the $\mathcal{O}(NMK)$ shared by every update for forming $\mathbf{A}=\mathbf{C}\mathbf{X}$ and $\mathbf{S}\mathbf{C}\mathbf{X}$, which for the gradient-based procedures is already the dominant term.}
    \label{tab:algorithms}
    \centering
    \resizebox{\columnwidth}{!}{%
    \begin{tabular}{lccc}
        \toprule
        \textbf{Optimization algorithm} & \textbf{Abbreviation} & \textbf{Computational Cost $\mathcal{O}_\mathbf{S}$} & \textbf{Computational Cost $\mathcal{O}_\mathbf{C}$} \\
        \midrule
        Nonnegative Least Squares~\cite{lawson1995solving,cutler1994archetypal} 
        & NNLS 
        & $\mathcal{O}_{\mathbf{S}}(NK^{3})$ 
        & $\mathcal{O}_{\mathbf{C}}(K|\mathcal{A}|^{3})$ \\ 
        Active Set \cite{chen:hal-00995911} & AS & {$\mathcal{O}_{\mathbf{S}}(KM + |\mathcal{A}|^2)$} & {$\mathcal{O}_{\mathbf{C}}(KM + |\mathcal{A}|^2)$}\\
        Principal Convex Hull Analysis \cite{morup2012archetypal} & PCHA & $\mathcal{O}_{\mathbf{S}}(NMK)$ & $\mathcal{O}_{\mathbf{C}}(NMK)$ \\
        Frank–Wolfe \cite{bauckhage2015archetypal} & FW & $\mathcal{O}_{\mathbf{S}}(NMK)$ & $\mathcal{O}_{\mathbf{C}}(NMK)$ \\
        Softmax + Adam \cite{wedenborg2024modeling, nakis2023characterizing} & SA & $\mathcal{O}_{\mathbf{S}}(NMK)$ & $\mathcal{O}_{\mathbf{C}}(NMK)$ \\
        Block Coordinate Descent \cite{behdin2024sparse} & BCD & $\mathcal{O}_{\mathbf{S}}(NMK)$ & $\mathcal{O}_{\mathbf{C}}(NMK)$ \\
        Multiplicative Update Rules \cite{seth2016probabilistic} & MUR & $\mathcal{O}_{\mathbf{S}}(NMK)$ & $\mathcal{O}_{\mathbf{C}}(NMK)$ \\
        Sequential Minimal Optimization \cite{wedenborg2025archetypal} & SMO & $\mathcal{O}_{\mathbf{S}}(NK^2)$ & $\mathcal{O}_{\mathbf{C}}(KN^2)$ \\
        Archetypoid Selection Algorithm \cite{vinue2015archetypoids} & ASA & $\mathcal{O}_{\mathbf{S}}(NK^{3})$ & $\mathcal{O}_{\mathbf{C}}\!\left(K(N-K)\,\mathcal{O}_{\mathbf{S}}\right)$ \\
        \bottomrule
    \end{tabular}
    }
\end{table}

\textbf{Active set (AS) procedure exploring sparsity.}
Notably, in \cite{chen:hal-00995911}, the inherent sparsity of the solution can be used within the active set procedure to significantly accelerate convergence. Specifically, the active set algorithm iteratively updates a subset $ \mathcal{A} $ of active (nonzero) variables. Given a current estimated $\mathbf{s}_n \in \Delta^{K-1}$ at a certain iteration, the algorithm defines a subset $\mathcal{A} = \{ j \; | \; a_j > 0 \}$, and searches for a direction $\mathbf{q} \in \mathbb{R}^K$ by solving the reduced problem:
\begin{align*}
\min_{\mathbf{q} \in \mathbb{R}^K} \ &\norm{ \mathbf{x}_n - (\mathbf{s}_n + \mathbf{q})\mathbf{Z}}_2^2 \
\text{s.t. } \sum_{k=1}^K q_k = 0, \ \ q_j = 0 \ \forall j \in \mathcal{A}^C,
\end{align*}
where $\mathcal{A}^C$ represents the complement of $\mathcal{A}$ in the index set $[K]$. The estimation is then updated as $\mathbf{s}_n = \mathbf{s}_n + \gamma \mathbf{q}$, with $\gamma$ adjusted to keep $\mathbf{s}_n$ within the feasible simplex $\Delta^{K-1}$. This active-set (AS) approach continues to refine the active set $\mathcal{A}$ until an optimal solution is achieved. Analogously, this method can be applied for optimizing with respect to $\mathbf{C}$.

\textbf{Sequential Minimal Optimization (SMO).}
In \cite{wedenborg2025archetypal} it was observed that when estimating $\mathbf{S}$ considering a limited number of archetypes, a very efficient and easily parallelizable inference procedure imposing the nonnegativity and sum-to-one constraint is the sequential minimal optimization (SMO) procedure originally used to solve for the support vector machine (SVM) quadratic optimization problem \cite{Platt1998SequentialMachines}. SMO considers the minimal quadratic subproblem formed by optimizing two elements in $\mathbf{s}_n$ at a time 
\begin{gather*}
\min_{s_{nk},s_{nk^\prime}}\|\mathbf{r}_n-\mathbf{c}_k\mathbf{X}s_{nk}-\mathbf{c}_{k^{\prime}}\mathbf{X}s_{nk^{\prime}}\|_2^2,\\
\mathbf{r}_n=\mathbf{x}_n-\sum_{k^{\prime\prime}\neq \{k,k^\prime\}}\mathbf{c}_{k^{\prime\prime}}\mathbf{X}s_{nk^{\prime\prime}},
\end{gather*}
which is equivalent to solving for $\alpha\in[0,1]$ the problem
\begin{gather*}
\min_{\alpha}\|\mathbf{r}_n-\mathbf{c}_k\mathbf{X}\alpha t-\mathbf{c}_{k^{\prime}}\mathbf{X}(1-\alpha)t\|_2^2,\\
t=s_{nk}+s_{nk^{\prime}},\ 
s_{nk}=\alpha t,\ 
s_{nk^\prime}=(1-\alpha) t.
\end{gather*}
By solving for $\alpha$ the sum of these two elements is redistributed between them, which results in a simple univariate second-order polynomial with closed-form solution; see also \cite{wedenborg2025archetypal} for further details. This procedure typically requires that all pairs be considered in order to converge sufficiently, producing a scaling of $\mathcal{O}_{\mathbf{S}}(NK^2)$ for the $\mathbf{S}$-update.

\textbf{Gradient-based approaches}
For scalable inference, gradient-based procedures can be used, yielding a per-iteration cost of $\mathcal{O}(NMK)$, typically at the cost of requiring substantially more iterations to convergence. In \cite{morup2012archetypal}, an alternating projected gradient approach for AA was proposed forming the principal convex hull analysis (PCHA) procedure. Considering the residual sum of squares given in Equation~\eqref{eq:AA_LS_objective}, the gradients with respect to $\mathbf{S}$ and $\mathbf{C}$ are 
\begin{align*}
\nabla_{\mathbf{S}} \operatorname{RSS} &= 2(\mathbf{SCX} \mathbf{X}^\transpose\mathbf{C}^\transpose - \mathbf{X} \mathbf{X}^\transpose\mathbf{C}^\transpose),
\\
\nabla_{\mathbf{C}} \operatorname{RSS} &= 2(\mathbf{S}^\transpose\mathbf{S}\mathbf{C}\mathbf{X}\mathbf{X}^\transpose - \mathbf{S}^\transpose\mathbf{X}\mathbf{X}^\transpose).
\end{align*}
By imposing the sum-to-one constraint through reparameterizing AA into the normalization invariant parameterizations $\tilde{\mathbf{s}}_n=\tfrac{\mathbf{s}_n}{\|\mathbf{s}_n\|_1}$ and $\tilde{\mathbf{c}}_k=\tfrac{\mathbf{c}_k}{\|\mathbf{c}_k\|_1}$ and using the chain rule of differentiation, the AA can be optimized using the following simple projected gradient descent procedure:
 \begin{align*}
 \mathbf{S} &\leftarrow \max\{\mathbf{0},\tilde{\mathbf{S}} - \eta_S (\nabla_{\mathbf{S}} E-\nabla_{\mathbf{S}} E\circ \tilde{\mathbf{S}}\mathbf{11}^\transpose)\},\ \tilde{\mathbf{s}}_n=\tfrac{\mathbf{s}_n}{\|\mathbf{s}_n\|_1},\\\mathbf{C} &\leftarrow\max\{\mathbf{0},\tilde{\mathbf{C}} - \eta_C (\nabla_{\mathbf{C}} E-\nabla_{\mathbf{C}} E\circ \mathbf{C}\mathbf{11}^\transpose)\},\ 
 \tilde{\mathbf{c}}_k=\tfrac{\mathbf{c}_k}{\|\mathbf{c}_k\|_1},
 \end{align*}
where $\max\{\mathbf{0},\mathbf{B}\}$ is evaluated elementwise and $\circ$ is the direct (elementwise) product $(\mathbf{A}\circ \mathbf{B})_{ij}=\mathbf{A}_{ij}\mathbf{B}_{ij}$. Notably, $\eta_S$ and $\eta_C$ are step size parameters that can be tuned by line search. In \cite{seth2016probabilistic} the sum-to-one constraint was instead imposed using the squared regularization penalty proposed in \cite{cutler1994archetypal} and $\mathbf{S}$ and $\mathbf{C}$ updated using multiplicative update rules (MUR) as originally proposed for nonnegative matrix factorization \cite{seung2001algorithms}. In \cite{EugsterPAMI} the conjugacy of the multinomial likelihood and Dirichlet distribution was further explored to derive closed form variational Bayesian (VB) inference updates. Furthermore, in \cite{hannachi2017archetypal} optimization using manifold approaches is leveraged. In \cite{bauckhage2015archetypal} efficient optimization via the Frank-Wolfe (FW) procedure \cite{frank_algorithm_1956} is also used to avoid complex quadratic programming when alternately updating $\mathbf{C}$ and $\mathbf{S}$. This procedure exploits the efficiency of computing the gradients $\nabla_{\mathbf{S}} E$ and $\nabla_{\mathbf{C}} E$. However, instead of the normalization invariant reparameterization the simplex constraint is achieved by performing subgradient updates along affine directions $\mathbf{e}_k - \mathbf{s}_n$ and $\mathbf{e}_n - \mathbf{c}_k$ within the simplices.
In \cite{behdin2024sparse} a block coordinate descent (BCD) projected gradient procedure was proposed using the efficient simplex projection operator described in \cite{duchi2008efficient}. Finally, softmax representation to impose the simplex constraint for efficient gradient descent has been considered in \cite{zouaoui_entropic_2023} and also used in combination with automatic differentiation and optimization using the Adam optimizer \cite{kingma2015adam} in the ordinal \cite{wedenborg2024modeling} and relational \cite{nakis2023characterizing} AA models.

\textbf{Archetypoid Selection Algorithm (ASA).}
When archetypes are required to correspond to actual observations, an alternative approach known as Archetypoid Selection Algorithm (ASA) was introduced in \cite{vinue2015archetypoids}.
In the ASA algorithm, the update of the matrix $\mathbf{C}$ is performed  using a greedy BUILD--SWAP strategy. For each candidate set of archetypoids, the coefficient matrix $\mathbf{S}$ is subsequently computed using an optimization method that allows $\mathbf{S}$ to be estimated, such as NNLS or PCHA.
The exchange that minimizes the objective function after updating $\mathbf{S}$ is selected as the new set of archetypoids, resulting in a complexity $\mathcal{O}_{\mathbf{C}}\!\left(K(N-K)\,\mathcal{O}_{\mathbf{S}}\right)$ per SWAP iteration.

\textbf{Implementations using surrogate AA objectives.}
In \cite{mei2018online} an algorithm based on online dictionary learning introduces a novel approach to AA by no longer defining the archetypes directly as convex combinations of the data observations. This is achieved through a spherically-constrained AA algorithm. Initially, the data points are projected onto the unit sphere, either by simple normalization or, in more complex cases, through stereographic projections. After projection, all points lie on the unit sphere. Here, the archetypes are defined as a matrix $\mathbf{A} \in \mathbb{R}^{K \times M}$ located within the unit sphere. To compute these archetypes, the optimization problem $\argmin_{\mathbf{A}, \mathbf{S}} \|\mathbf{X} - \mathbf{S}\mathbf{A}\|_\text{F}^2 
$ is solved subject to $\|\mathbf{a}_k\|_\text{2}^2 \leq 1, \quad \mathbf{s}_n \in \Delta^{K-1}$. Since the archetypes are constrained to reside within the unit sphere, their convex combinations form a convex hull within the sphere. Given that the data points are also normalized to lie on the unit sphere, they are guaranteed to fall outside this convex hull. In the learning phase, the algorithm iteratively adjusts the convex hull surfaces to minimize the overall representation error across the dataset, effectively pushing the surfaces closer to the data. Solving this optimization problem produces archetypes that approximately correspond to the extreme points of the dataset. 

An algorithm proposed in \cite{abrol2020geometric} leverages the underlying geometry and sparsity patterns of convex representations to identify archetypes on the convex hull of the data. Instead of optimizing $\mathbf{S}$ and $\mathbf{C}$ directly with respect to the entire data matrix $\mathbf{X}$, the authors define an alternative objective function that enables efficient and independent learning of $\mathbf{C}$ without requiring direct interaction with $\mathbf{X}$. This is achieved by learning~$\mathbf{C}$ in the coefficient space rather than the signal space. The method begins by solving the following optimization problem $\argmin_{\mathbf{Z}} \|\mathbf{X} - \mathbf{Z}\mathbf{X}\|_\text{F}^2$ subject to 
\mbox{$\operatorname{diag}(\mathbf{Z}) = 0$} removing one degree of freedom such that $\mathbf{z}_n \in \Delta^{N-2}$. Once $\mathbf{Z}$ is obtained, the coefficient matrices $\mathbf{S}$ and~$\mathbf{C}$ can be alternately derived by solving $\argmin_{\mathbf{S}, \mathbf{C}} \|\mathbf{Z} - \mathbf{S}\mathbf{C}\|_\text{F}^2$ subject to $\mathbf{s}_n \in \Delta^{K-1}$ and $\mathbf{C}_k \in \Delta^{N-1}$.

AA approximations using quantum annealing based on surrogate objectives that form quadratic unconstrained binary optimization problems were considered in \cite{feld_approximating_2020}. More recently, \cite{han2022probabilistic} presented a data compression technique based on a randomized Krylov subspace method to decrease data dimensionality. This method minimizes the need for frequent queries to high-dimensional datasets and represents a novel approach within the context of AA. Furthermore, the study proposes using random projections to approximate the data's convex hull, preserving distances approximately, and thereby reducing the dictionary's cardinality for archetype representation. Both techniques enhance the efficiency of the AA algorithm by reducing the dataset's size and dimensionality.
A related idea is to apply a PCA on the data matrix $\mathbf{X}$ for dimensionality reduction before running AA \cite{richardson2021identifying,risbey2021identification,gimbernat2022archetypal}. This approach was coined reduced space archetypal analysis (RSAA) and shown to yield results similar to vanilla AA \cite{black2022archetypal}.

Importantly, surrogate models can provide an efficient approach to handle the high computational demands of AA, especially with large datasets. Approximating the model's behavior reduces the analysis time, enabling quick and affordable exploration of data patterns.

\textbf{Practical guidance}. The complexities reported in Table~\ref{tab:algorithms} are per-iteration costs, and should be read together with the number of iterations each procedure requires. For small and moderate $N$ with a low number of archetypes $K$, the QP formulation solved by NNLS~\cite{cutler1994archetypal} solves each alternating subproblem directly rather than iteratively, but its cubic dependence on $|\mathcal{A}|$ and $K$ makes it impractical as either grows; the
Active Set procedure of \cite{chen:hal-00995911} is preferable whenever the rows of $\mathbf{S}$ are strongly sparse, since the cost is then governed by $|\mathcal{A}|\ll K$, and SMO~\cite{wedenborg2025archetypal} reduces the $\mathbf{S}$-update to $\mathcal{O}_{\mathbf{S}}(NK^2)$. Gradient-based procedures such as PCHA~\cite{morup2012archetypal}, FW~\cite{bauckhage2015archetypal} and BCD~\cite{behdin2024sparse} scale linearly in $N$, $M$ and $K$ and are therefore the natural default for large datasets, at the price of requiring substantially more iterations; among them, PCHA with line search is a robust general-purpose choice, while FW and BCD differ mainly in how the simplex constraint is enforced. MUR~\cite{seth2016probabilistic} and the variational updates of \cite{EugsterPAMI} are motivated by probabilistic formulations with non-Gaussian likelihoods rather than by speed, while the softmax reparameterization combined with automatic differentiation and Adam~\cite{zouaoui_entropic_2023,wedenborg2024modeling,nakis2023characterizing} is the appropriate choice whenever AA is embedded in a larger differentiable model, or when GPU training is required. ASA~\cite{vinue2015archetypoids} and its large-data extension~\cite{Vinue21} should only be used when the archetypes are required to be actual observations, either because averaging is not meaningful for the data at hand or because domain experts need to inspect real cases. Finally, the surrogate and dimensionality-reduction approaches discussed above \cite{mei2018online,abrol2020geometric,han2022probabilistic,black2022archetypal} are worth the loss of exactness only when the original problem is intractable at the scale considered, and the same reasoning applies to robust, sparse and deep extensions of AA, whose additional hyperparameters and computational cost are justified only when the corresponding assumption is violated: heavy contamination by outliers, high-dimensional data requiring feature-level interpretability, or data lying on a strongly nonlinear manifold for which a convex hull spanned by convex combinations of observations is a poor approximation.

\subsection{Strategies for initialization of archetypes}\label{sub:strategies_for_archetypes_initialization}

An important consideration when fitting AA models is the initialization of archetypes. Proper initialization is crucial for improving the algorithm's convergence speed and ensuring the reliability of the resulting archetypes.

In \cite{cutler1994archetypal}, a random initialization method is proposed. In this approach, archetypes are selected from the dataset using a uniform distribution. Specifically, each index $n^{\text{next}} \sim \text{Uniform}([N])$ is randomly chosen, where $[N]$ denotes the set of indices of the dataset.
\looseness=-1

An alternative approach is the FurthestSum algorithm, introduced by \cite{morup2010archetypal, morup2012archetypal}, which adapts the FurthestFirst \cite{hochbaum1985best} algorithm devised for the $k$-center problem into an initialization approach suitable for AA. In this method, the first archetype is selected randomly, just as in the original FurthestFirst approach. However, subsequent archetypes are chosen based on the furthest distance from the already selected archetypes. The next archetype index is selected according to the rule
\begin{align*}n^{\text{next}} = \argmax_{n \in [N]} \left( \sum_{\mathbf{a} \in \mathbf{A}} \norm{\mathbf{x}_n - \mathbf{a}}_2 \right),
\end{align*}
where $\mathbf{A}$ represents the set of already selected archetypes and $\mathbf{x}_n$ is a data point from the dataset. This method guarantees that each new archetype is as far as possible from the existing ones and thus residing on the convex hull, at the same time enhancing the diversity and representativeness of the archetypes across the dataset.

A recent initialization procedure is AA++ \cite{mair2024archetypal}. It is inspired by the $k$-means++ initialization \cite{arthur2007kmeans} and the idea of AA++ is to randomly select the first archetype while the following archetypes are chosen based on a probability distribution $P(n) \propto \min_{\mathbf{a} \in \operatorname{conv}(\mathbf{A})} \norm{\mathbf{x}_n - \mathbf{a}}_2^2$, which is proportional to the distance between a data point and the convex hull of the already chosen archetypes.
AA++ ensures that points farther from existing archetypes have a higher probability of being selected, thereby promoting better coverage of the dataset and a faster convergence of AA.
Note that AA++ can be approximated by $k$-means++ \cite{mair2024archetypal}.

\begin{figure*}[!ht]
\centering
\includegraphics[width=0.8\linewidth]{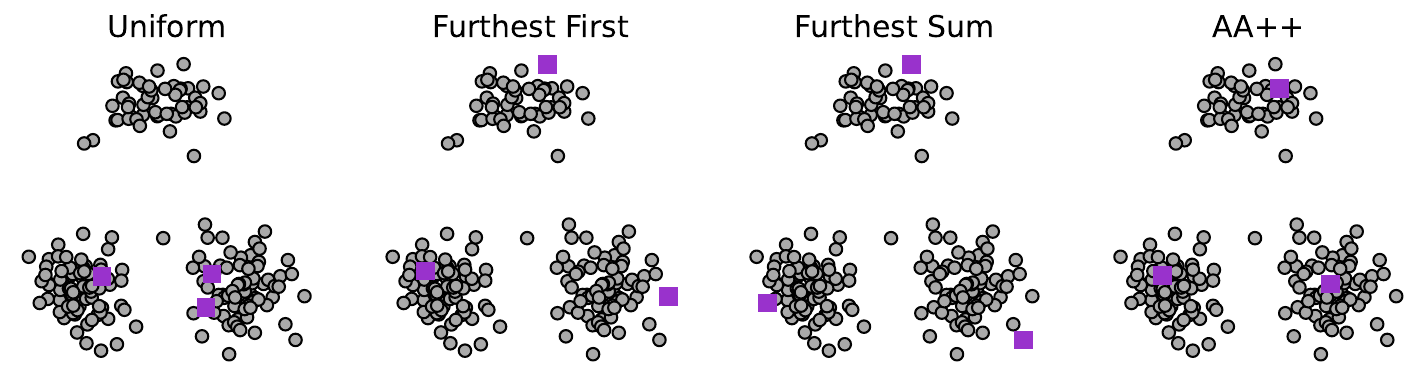}
\caption{A comparison of the following initialization techniques of AA for $K=3$: Uniform, FurthestFirst, FurthestSum, and AA++.}
\label{fig:AAinits}
\end{figure*}

Another line of work directly explores the use of initialization techniques originally designed for clustering. For instance, strategies such as $k$-means++ \cite{arthur2007kmeans}, FurthestFirst \cite{hochbaum1985best}, and the Incremental Anomalous Pattern (AP) algorithm \cite{nascimento2019unsupervised} can be adapted for AA. Although these methods are not explicitly designed for AA, unlike those above, they can still effectively initialize the prototypes as shown in \cite{mair2024archetypal}. Notably, the AP algorithm offers the added advantage of determining the number of prototypes to use, making it particularly valuable in scenarios where the optimal number of archetypes is unknown.

It is important to note that all these initialization methods may introduce redundant archetypes. However, in \cite{suleman_ill-conceived_2017}, an algorithm is introduced to recycle the archetypes by removing redundant archetypes. Examples of these initializations can be seen in Figure~\ref{fig:AAinits}, and a recent, detailed comparison of many initialization approaches can be found in \cite{mair2024archetypal}.

\subsection{Dataset size reduction techniques}

Reducing dataset size is crucial for improving computational efficiency in AA, particularly when dealing with large datasets. Techniques that reduce dataset size while preserving its core structure and information allow analyses to be performed more quickly without significantly affecting the quality of the results.

One common approach is to calculate a subset, $\bar{\mathbf{X}}$, of the dataset and use it to compute the archetypes. Since $\bar{\mathbf{X}}$ is smaller than the original dataset, the algorithm converges more quickly.

For example, in \cite{thurau2009convex, bauckhage2009making, thurau2011convex}, the archetypes are computed from a subsample of points on the convex hull of $\mathbf{X}$, which is constructed by taking the union of points found on the convex hulls of different 2D projections of $\mathbf{X}$.

Alternatively, in \cite{mair2017frame}, $\bar{\mathbf{X}}$ is set to be the subset of the dataset that resides on the boundary of the convex hull of the dataset, the so-called frame. Note that the frame can also be computed when using kernels \cite{mair2018frame}, allowing its usage for kernel AA.

In another approach, presented in \cite{mair2019coresets}, a representative subset~$\bar{\mathbf{X}}$ is generated by sampling points from a distribution $P(n) \propto \|\mathbf{x}_n - \bm{\mu}\|_2^2$, where $\bm{\mu}$ is the mean of the data points.
This subset is a coreset, and theoretical guarantees on the approximation error can be stated.
It is noteworthy that the points in the subset are weighted, and a weighted version of AA must be used when learning the model.

On the other hand, there is a technique for reducing dataset size that is applied at each optimization step, rather than before the algorithm starts. This method, introduced in \cite{bauckhage2009making}, is based on the observation that data points within the convex hull of the archetypes do not contribute to the residual minimized by the archetype algorithm. As a result, in each iteration, the dataset is divided into two sets: $\mathbf{X}^-$ and $\mathbf{X}^+$. The set $\mathbf{X}^-$ contains points that can be exactly represented as convex combinations of archetypes, while $\mathbf{X}^+$ contains points that can only be approximated. By focusing only on $\mathbf{X}^+$, which is smaller than the full dataset, the algorithm converges more quickly. Moreover, as archetypal estimates improve over iterations, the number of points outside the convex hull decreases, further reducing the optimization problem size and accelerating the algorithm. Notably, this approach is relevant in the low-dimensional setting where the convex set forming the convex hull can be assumed small. As the convex set grows exponentially with dimensionality (see also \cite{morup2012archetypal}) this approach is not suitable for high-dimensional problems.
\looseness=-1

\subsection{Assessing model robustness}
The quality of AA solutions can be assessed in terms of goodness of fit and robustness.  
Assessing the robustness of the model is essential to ensure that the archetypes and the insights they provide are reliable and meaningful. Several techniques can be used to evaluate the stability and consistency of the model, each addressing different aspects of the analysis.

A critical consideration in AA is the use of random initializations and error bars on the loss. AA typically involves an optimization process that starts from random initial points in the parameter space. The randomness of these initializations can lead to different local optima, potentially affecting the resulting archetypes. To evaluate the stability of the results, multiple AA runs with different initializations can be performed, and the loss (e.g., reconstruction error) can be monitored across them. By visualizing error bars on the loss values, we can gauge the variability of the results. Smaller error bars indicate that the model is stable, while larger error bars suggest greater sensitivity to initial conditions. 

However, results may still produce reliable estimates of error whereas the underlying extracted components substantially change. To assess the consistency of the extracted $\mathbf{S}$ matrices normalized mutual information between two estimated $\mathbf{S}$ and $\mathbf{S}^\prime$ matrices was proposed in \cite{hinrich2016archetypal} exploiting that $\mathbf{P}=\frac{1}{N}\mathbf{S}^\transpose\mathbf{S}^{\prime}$ forms a joint distribution with elements $p_{k,k^\prime}$ for which the mutual information can be quantified by $\text{MI}(\mathbf{S},\mathbf{S}^\prime)=\sum_{k,k^\prime}p_{k,k^\prime}\log{\frac{p_{k,k^\prime}}{\sum_{k^{\prime\prime}}p_{k,k^{\prime\prime}}\cdot \sum_{k^{\prime\prime}}p_{k^{\prime\prime},k^\prime}}}$ and normalized to a similarity score between 0 and 1, i.e.,
\begin{equation*}
\text{NMI}(\mathbf{S},\mathbf{S}^\prime)=\frac{2\text{MI}(\mathbf{S},\mathbf{S}^\prime)}{\text{MI}(\mathbf{S},\mathbf{S})+\text{MI}(\mathbf{S}^\prime,\mathbf{S}^\prime)}.
\end{equation*}
For the consistency of archetypes 
$\mathbf{A}=\mathbf{C}\mathbf{X}$ for two matrices $\mathbf{A}$ and $\mathbf{A}^\prime$ greedily matching the components by their proximity to each other and averaging these distances to an average squared distance $\bar{d}^2$ can be used to define a similarity score as 
\begin{equation*}
\text{sim}(\mathbf{A},\mathbf{A}^\prime)=1-\bar{d}^2/\bar{\sigma}^2_{\mathbf{X}},
\end{equation*}
where $\bar{\sigma}^2_{\mathbf{X}}$ is the variance summed across the $M$ features of $\mathbf{X}$.
\looseness=-1

Bootstrapping is another technique used to assess the robustness of the archetypes by generating multiple resamples of the original data with replacement \cite{hart2015inferring}. AA is applied to each of these resampled datasets, and the resulting archetypes are compared to measure their variability. Bootstrapping helps determine whether the observed archetypes are consistent across different subsets of the data and whether they generalize to new, unseen data.

Using the above techniques, we can robustly assess the reliability of the archetypes produced by AA. This ensures that the insights drawn from the analysis are both meaningful and reproducible, making the model more reliable across diverse applications.
\looseness=-1

\subsection{Validating the selection of the number of archetypes}

Strategies for selecting the number of archetypes, i.e., $K$, have traditionally relied on visual inspection. Common validation methods include scree plot or elbow criteria, which analyze the monotonic behavior of either an objective function \cite{eugster2009spider} or a measure of the variation explained by different models \cite{cutler1994archetypal, morup2012archetypal} while others use BIC-based model selection criteria \cite{prabhakaran2012automatic}. More recently, \cite{suleman_validation_2017} introduced an information-theoretic criterion specifically adapted to assess the goodness-of-fit in AA. This metric, analogous to the AIC-like measure discussed in \cite{suleman_measuring_2017}, is defined as 
$$v_{\text{AA}}(K) = \ln{\left( \frac{1}{N \cdot M} \|\mathbf{X} - \tilde{\mathbf{X}}\|_\text{F}^2 \right)} + 2 \frac{2K - 1}{\text{tr}\left({\Sigma_{\tilde{\mathbf{X}}} \Sigma_{\mathbf{X}}^{-1}}\right)},$$
where $\Sigma_{\tilde{\mathbf{X}}}$ and $\Sigma_{\mathbf{X}}$ represent the covariance matrices of $\tilde{\mathbf{X}}$ and $\mathbf{X}$, respectively.

\subsection{Accessible tools and software}

Software tools for AA are essential for researchers aiming to apply these methods effectively. Accessible software makes AA practical for more fields, supporting reproducible research and encouraging broader adoption across diverse disciplines. 

Table~\ref{tab:software} offers an overview of the main software tools for AA. Each column in the table is intended to help readers quickly identify which package best fits their methodological and computational needs. \textit{Name} identifies the tool or software package, while \textit{Language} specifies the programming environment in which it is implemented. \textit{Description} provides a concise overview of the tool's functionality, and \textit{Models} indicates the AA variants supported. The \textit{Optimization algorithms} column details the computational methods used and allows each software tool to be directly related to the algorithms introduced in Section~\ref{sec:advancemenents2AA}. \textit{Sparsity} denotes whether the tool supports sparse data or not, \textit{Nonlinearity} indicates the availability of nonlinear extensions, \textit{Robustness} reflects the tool’s ability to handle outliers or noisy data, and \textit{Weighted} specifies whether observation or feature weighting is supported.

Together, these columns offer a structured comparison that helps users select a package aligned with their data type, computational constraints, and modeling goals.

\begin{table*}[ht]
\caption{Summary of the main available packages and tools for archetypal analysis.}
\label{tab:software}
\centering
\resizebox{\textwidth}{!}{% 
\begin{tabular}{p{0.18\textwidth}p{0.12\textwidth}p{0.3\textwidth}p{0.12\textwidth}p{0.2\textwidth}cccc}
\toprule
\textbf{Name} & \textbf{Language} & \textbf{Description} & \textbf{Models} &  \textbf{Optimization algorithms} & \textbf{Sparsity} & \textbf{Nonlinearity} & \textbf{Robustness} & \textbf{Weighted} \\

\midrule
\multicolumn{9}{c}{\textsc{LEAST-SQUARES AA}}\\

\href{https://cran.r-project.org/web/packages/archetypes/index.html}{\texttt{archetypes}} \cite{eugster2009spider} & R & 
Package for AA & AA & NNLS \cite{cutler1994archetypal} & \ding{55} & \ding{55} & \ding{51} & \ding{51} \\
\href{https://cran.r-project.org/web/packages/adamethods/index.html}{\texttt{adamethods}} \cite{vinueR,Vinue21} & R &
Package for archetypoid analysis & ADA & NNLS \cite{cutler1994archetypal}, ASA \cite{vinue2015archetypoids} & \ding{55} & \ding{55} & \ding{51} & \ding{55} \\
\href{https://archetypes.readthedocs.io/}{\texttt{archetypes}} \cite{alcacerieee24} & Python & Package for AA and visualization tools & AA, ADA, BiAA & PCHA \cite{morup2012archetypal}, NNLS \cite{cutler1994archetypal}, PAA \cite{seth2016probabilistic} & \ding{55} & \ding{51} & \ding{55} & \ding{51} \\
\href{https://thoth.inrialpes.fr/people/mairal/spams/index.html}{\texttt{SPAMS}} \cite{chen:hal-00995911} & 
R, Matlab, Python & 
Sparse modeling package including AA  & AA & AS \cite{chen:hal-00995911} & \ding{55} & \ding{55} & \ding{55} & \ding{55} \\
\href{https://vitkl.github.io/ParetoTI/}{\texttt{ParetoTI}} \cite{hart2015inferring} & R & Package for AA and Pareto task inference & AA & PCHA \cite{morup2012archetypal} & \ding{55} & \ding{55} & \ding{55} & \ding{55} \\

\href{https://github.com/inria-thoth/HySUPP}{\texttt{HySUPP}} \cite{rasti2024image} & Python & Hyperspectral unmixing package including AA & AA & AS \cite{chen:hal-00995911} & \ding{55} & \ding{55} & \ding{55} & \ding{55} \\

 \href{https://mortenmorup.dk/?page_id=2}{PCHA} 
 \cite{morup2010archetypal,morup2012archetypal} & 
Matlab & 
AA for regular data, kernels, and sparse data & AA & PCHA \cite{morup2012archetypal} & \ding{51} & \ding{51} & \ding{55} & \ding{55} \\

\href{https://github.com/ulfaslak/py\_pcha}{py\_pcha} \cite{morup2010archetypal} & Python & Fast AA using principal convex hull analysis & AA & PCHA \cite{morup2012archetypal}  & \ding{55} & \ding{55} & \ding{55} & \ding{55} \\

\href{https://github.com/AI-sandbox/archetypal-analysis}{archetypal-analysis} \cite{gimbernat2022archetypal} & Python & AA for genomic data & AA & NNLS \cite{cutler1994archetypal} & \ding{55} & \ding{55} & \ding{55} & \ding{55} \\

\href{https://github.com/atmguille/archetypal-analysis}{archetypal-analysis} & Python, R & PCHA and Frank-Wolfe based AA & AA & NNLS \cite{cutler1994archetypal}, PCHA \cite{morup2012archetypal}, FW \cite{bauckhage2015archetypal} & \ding{55} & \ding{55} & \ding{55} & \ding{55} \\

\href{https://github.com/kayhanbehdin/SparseAA}{SparseAA} \cite{behdin2024sparse} & Julia & Sparse AA & AA & BCD \cite{behdin2024sparse} & \ding{55} & \ding{55} & \ding{51} & \ding{55} \\

\href{https://github.com/BehnoodRasti/SUnAA}{SUnAA} \cite{rasti2023sunaa} & Python & Sparse Unmixing using AA & AA & AS \cite{chen:hal-00995911}  & \ding{55} & \ding{55} & \ding{55} & \ding{55} \\

\href{https://compbio.mit.edu/ACTION/}{ACTION} \cite{mohammadi2018geometric} & C, C++, Matlab, R & AA for Single-cell transcriptomic data & AA & PCHA \cite{morup2012archetypal} & \ding{55} & \ding{55} & \ding{55} & \ding{55} \\

\midrule
\multicolumn{9}{c}{\textsc{AA BASED ON DIFFERENT LIKELIHOOD SPECIFICATIONS}} \\

\href{https://github.com/aalab/paa}{PAA} \cite{seth2016probabilistic} & Matlab, R & Probabilistic AA & AA & NNLS \cite{cutler1994archetypal}, MUR \cite{seth2016probabilistic} & \ding{55} & \ding{55} & \ding{55} & \ding{55} \\

\href{https://github.com/aalab/naa}{NAA} \cite{EugsterPAMI} & Matlab, C & Nominal AA & AA & VB \cite{EugsterPAMI} & \ding{55} & \ding{55} & \ding{55} & \ding{55} \\

\href{https://github.com/ChristianDjurhuus/RAA}{RAA} \cite{nakis2023characterizing} & Python & Relational AA & AA & SA \cite{nakis2023characterizing} & \ding{51} & \ding{55} & \ding{55} & \ding{55}  \\

\href{https://github.com/anders-s-olsen/DirectionalArchetypalAnalysis}{DAA} \cite{olsen2022combining} & Matlab & Directional AA & AA & PCHA \cite{olsen2022combining} & \ding{55} & \ding{55} & \ding{55} & \ding{55} \\

\href{https://github.com/Maplewarrior/OrdinalArchetypalAnalysis}{OAA} \cite{wedenborg2024modeling}  & Python & Ordinal AA & AA & SA \cite{wedenborg2024modeling} & \ding{55} & \ding{55} & \ding{55} & \ding{55} \\

\midrule
\multicolumn{9}{c}{\textsc{NONLINEAR DEEP AA}}
\\

\href{https://github.com/KrishnaswamyLab/AAnet}{AAnet} \cite{van2019finding} & Python & Scalable AA of large and nonlinear datasets & AA & Neural Network \cite{van2019finding} & \ding{55} & \ding{51} & \ding{55} & \ding{55}  \\

\href{https://github.com/AprilYuge/scAAnet}{scAAnet} \cite{wang2022non} & Python & Single-cell AA neural network & AA & Neural Network \cite{wang2022non} & \ding{55} & \ding{51} & \ding{55} & \ding{55} \\

\href{https://github.com/bmda-unibas/DeepArchetypeAnalysis}{DeepArchetypeAnalysis} \cite{keller2019deep} & Python & Deep AA & AA & Neural Network \cite{keller2019deep} & \ding{55} & \ding{51} & \ding{55} & \ding{55} \\

\bottomrule
\end{tabular}% 
} % resizebox
\end{table*}

\subsection{Results visualization}

In any type of machine learning analysis, interpreting the results is a crucial step. In the case of AA, this can primarily be approached in two ways: by analyzing the scores, which measure how similar each observation is to each archetype, or by directly examining the archetypes themselves. When focusing on the archetypes $\mathbf{A}=\mathbf{CX}$, we can assess their values in terms of the percentiles within the distribution of each variable. Several types of visualizations, where archetypes are consistently represented by colors, can aid in this process. These visualizations can be easily generated using the Python package \texttt{archetypes} described in Table~\ref{tab:software}, which includes convenient tools for creating and customizing plots.

\begin{figure}[t]
    \centering
    \includegraphics[width=\linewidth]
    {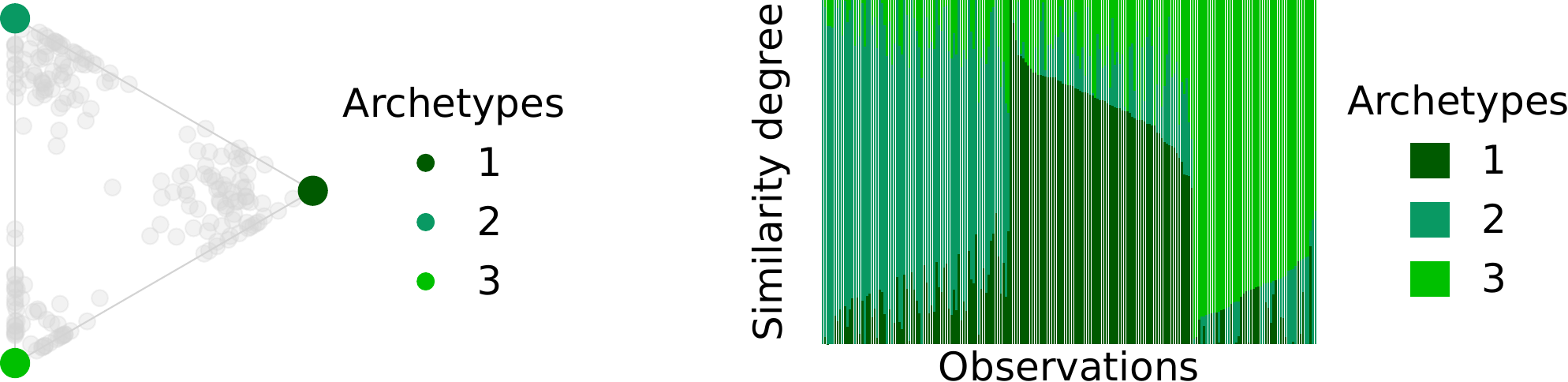}
    \caption{Left panel: Simplex plot representing the scores, where each observation is a point, and the archetypes are positioned at the colored vertices. The location of each point indicates its relative contribution from the archetypes. Right panel: Stacked bar chart of the scores, where each bar represents an observation and the colored segments correspond to the contributions of each archetype.}
    \label{fig:simplexstacked_bar}
\end{figure}

Figure~\ref{fig:simplexstacked_bar} illustrates visualizations of $\mathbf{S}$ defining how each observation can be visualized as a mixture of the archetypes. The left panel of the figure illustrates a simplex plot, where each point corresponds to an observation. The position of each point reflects the relative contribution of the archetypes, represented by colored vertices. This type of plot is particularly effective for identifying clusters of observations that share similar contributions from the archetypes. The simplex plot works well for up to three archetypes when visualized in 2D, but it becomes redundant in how points are reconstructed when including more archetypes.

As an alternative to the simplex plot, the right panel of Figure~\ref{fig:simplexstacked_bar} displays a stacked bar chart of the scores $\mathbf{S}$. Each bar corresponds to an observation, segmented into colored sections that indicate the contribution of each archetype. This chart provides a straightforward way to compare how different archetypes combine across the observations and accurately represents $\mathbf{S}$ for arbitrary numbers of archetypes.

In general, the best way to visualize the AA results is typically very problem specific. For instance, when working with geospatial data or 3D shapes, other types of visualizations tailored to these data structures are typically more appropriate. For example, heatmaps or choropleth maps might be useful for representing archetypes in spatial datasets, while 3D scatter plots or mesh visualizations could be used to analyze shapes \cite{plos20}.
\looseness=-1

\section{Prominent applications of AA}\label{sec:ProminentApplicationsofAA}
Below we outline prominent applications of AA within disparate fields of science.  Here, we provide an overview of how AA is applied across different fields, while detailed application information is presented in the supplemental material. Figure~\ref{fig:AA_appli} summarizes the AA application areas.
In Figures~\ref{fig:AAexampleFinches}--\ref{fig:AAexampleCongress}, we highlight five example analyses of AA covering biological, chemical, remote sensing, outlier detection, and social science applications. 
In the analyses the data has been centered by subtracting the mean.
\looseness=-1

\begin{figure}[t]
\centering
\includegraphics[width=\columnwidth]{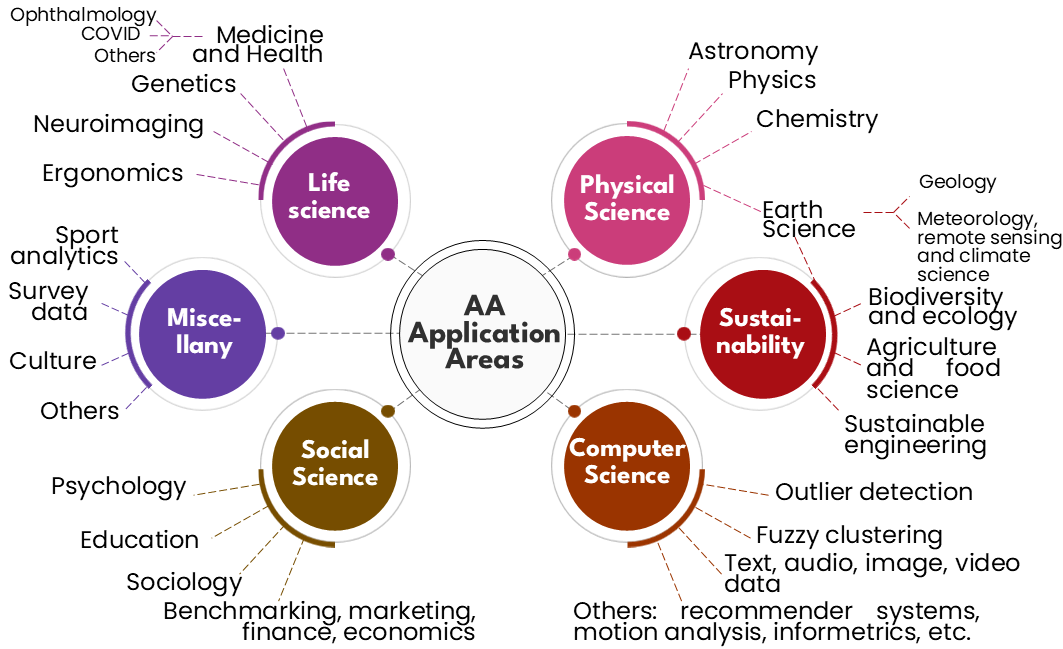}
\caption{Bubble map with the main areas of AA applications.
}
\label{fig:AA_appli}
\end{figure}

\subsection{Life science}

AA has found wide applications within the life-sciences since it provides an intuitive geometric framework for identifying extreme, biologically meaningful profiles (the archetypes). This allows researchers to capture trade-offs, uncover latent structure, and describe complex biological or clinical systems in an interpretable way for guiding scientific or clinical decision-making.
\looseness=-1

\begin{figure}[th]
\centering
\includegraphics[width=\columnwidth]{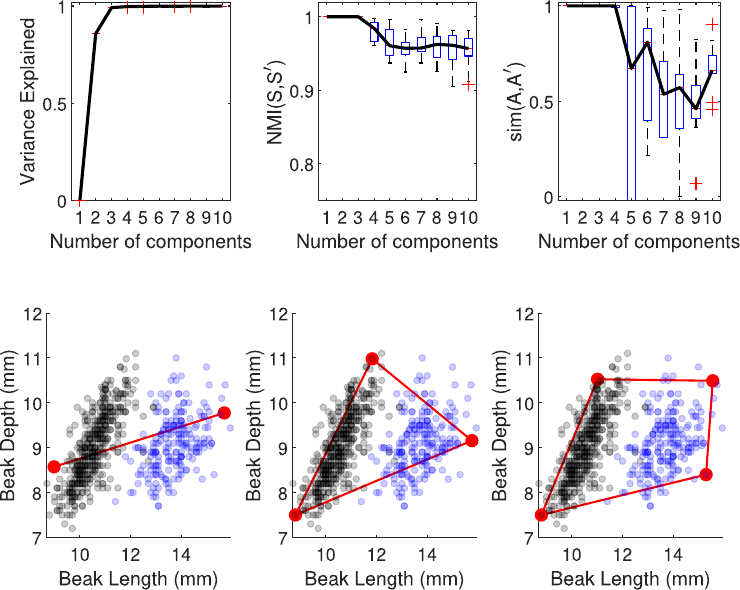}
\caption{Analysis of data on Finches\protect\footnotemark in which two features are measured based on beak length and depth for scandens (blue dots) and fortis (black dots) finches illustrated for a $K=2$, $K=3$, and $K=4$ AA model. When inspecting the variance explained we observe substantial improvements up until $K=3$, and reproducibility as measured using $\text{NMI}(\mathbf{S},\mathbf{S}^\prime)$ and $\text{sim}(\mathbf{A},\mathbf{A}^\prime)$ we observe that a $K=3$ model is also reliably recovering $\mathbf{S}$ and $\mathbf{A}$. Whereas the $K=2$ AA model mainly discriminates between the two types of finches, the $K=3$ characterizes in particular variability in the fortis finch class whereas the $K=4$ further subdivides the scandens finch class.
}
\label{fig:AAexampleFinches}
\end{figure}

In evolutionary biology, AA has been used to extract polytopes describing evolutionary trade-offs \cite{shoval2012evolutionary} in traits, where archetypes represent optimal strategies across tasks (see Figure~\ref{fig:AAexampleFinches}).  AA has been applied in genetics \cite{Morup2013} to characterize population-level genetic profiles and trait expression patterns. Single-cell RNA-seq studies adopt AA (including kernel and deep variants \cite{wang2022non}) to model cell-type diversity through interpretable extreme transcriptional states. AA is also used in clinical genomics and transplant medicine \cite{reeve2017assessing}.

\footnotetext{Data taken from \url{https://datadryad.org/dataset/doi:10.5061/dryad.g6g3h}}

In neuroimaging, AA summarizes high-dimensional measured functional  brain activity \cite{hinrich2016archetypal} in terms of distinct activation patterns. In epidemiology, AA and ADA are applied to COVID-19 data \cite{grane2022looking} to capture distinct spatio-temporal outbreak archetypes. In ophthalmology, AA has facilitated characterization of visual field loss patterns \cite{elze2015patterns}, whereas stacked AA has been used to extract clinically meaningful extreme visual-field patterns in \cite{mahotra2020patterns}. AA and ADA have also been applied to ergonomics  \cite{EpiVinAle} to identify boundary body shapes for design requirements. Furthermore, within life science, several AA extensions such as kernel, deep AA, ADA, hierarchical, and multimodal formulations, are used to capture nonlinear structure better, integrate heterogeneous data sources, and provide more flexible yet still interpretable representations of biological and clinical variability.
\looseness=-1

\subsection{Physics and chemistry}

\begin{figure}[th]
\centering
\includegraphics[width=\columnwidth]{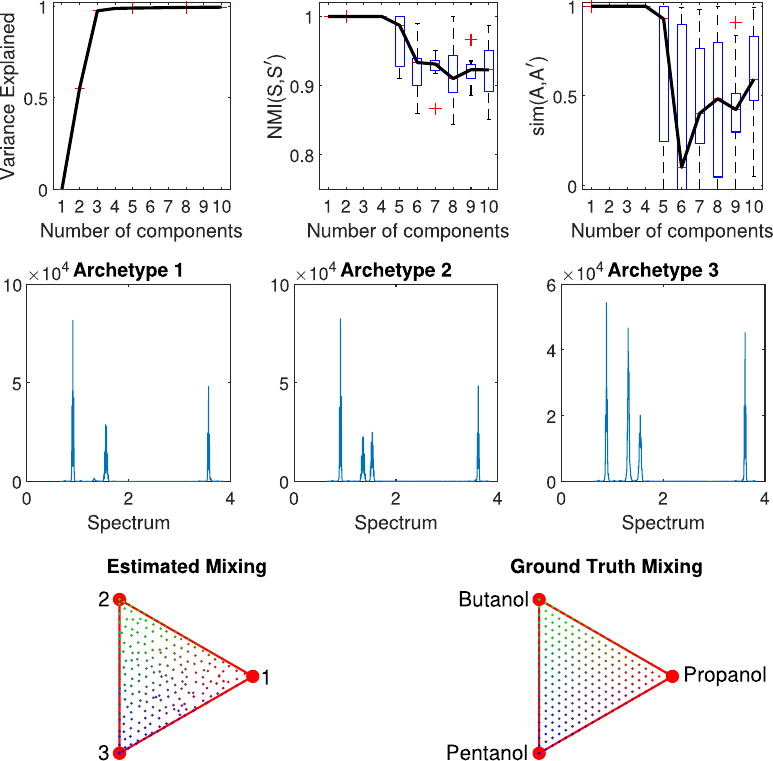}
\caption{Analysis of data on mixture designed NMR experiment containing systematically different fractions of propanol, butanol and pentanol\protect\footnotemark. When inspecting the variance explained we observe substantial improvements up until $K=3$ and reproducibility as measured using $\text{NMI}(\mathbf{S},\mathbf{S}^\prime)$ and $\text{sim}(\mathbf{A},\mathbf{A}^\prime)$ we observe that a $K=3$ reliably recovers $\mathbf{S}$ and $\mathbf{A}$. This three component model well corresponds to the three compounds whereas the estimated concentration fractions (given at the bottom) well correspond to the true concentration levels of the three compounds in each sample.
}
\label{fig:AAexampleNMR}
\end{figure}

AA has been broadly adopted in physics and chemistry because it provides an interpretable, extremal-point-based decomposition of complex data. The central motivation for using AA in these domains is that many physical and chemical measurements can be viewed as mixtures of a small number of fundamental, ``pure'' sources that correspond to meaningful physical or chemical constituents. Applications range from astrophysical data, where galaxy spectra are decomposed into emissions from different stellar and nebular populations \cite{chan2003archetypal}, to various dynamical systems \cite{stone1996archetypal}. In chemistry, AA has been used to analyze NMR \cite{morup2012archetypal} (see Figure~\ref{fig:AAexampleNMR}) and infrared spectra,  to characterize virtual nanomaterial samples \cite{fernandez2015identification}. 

\footnotetext{Data taken from \url{https://ucphchemometrics.com/datasets/}}

\subsection{Climate science and sustainability}

AA is widely used in climate science and computational sustainability because it provides an interpretable framework for identifying extreme, physically meaningful patterns within high-dimensional environmental, ecological, or remote sensing data. Many climate and ecological phenomena can be viewed as mixtures of a few characteristic states, enabling both dimensionality reduction and physically interpretable pattern discovery.
\looseness=-1

In climate applications, AA has been used to characterize precipitation patterns \cite{steinschneider2015daily} and other climate and environmental patterns.  Beyond climate, the same motivation drives AA's use in computational sustainability \cite{Thurau12}. 

\begin{figure}[t]
\centering
\includegraphics[width=\columnwidth]{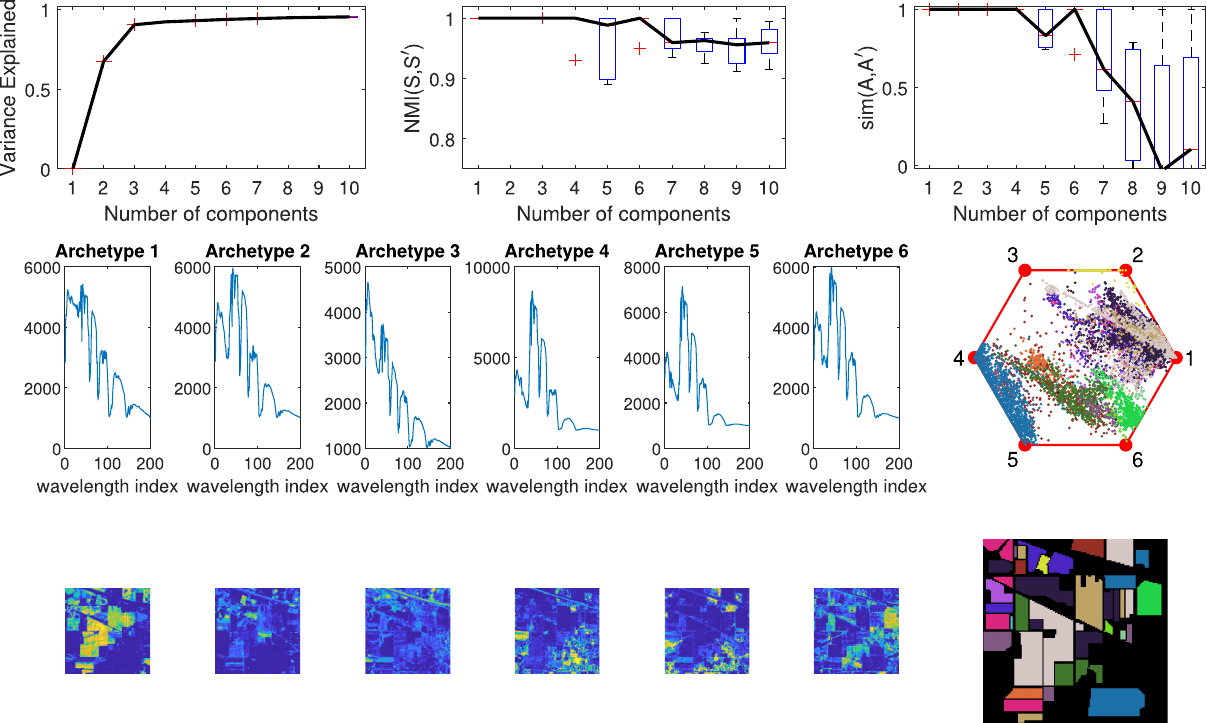} 
\caption{Analysis of a remote sensing hyperspectral image data\protect\footnotemark in which an image of a landscape in Indiana, US is measured across 200 wavelengths (removing bands pertaining to water absorption) in a $145 \times 145$ image. The image is turned into a matrix of $200$ wavelength features by $145^2$ pixel observations. When inspecting the variance explained and reproducibility as measured using $\text{NMI}(\mathbf{S},\mathbf{S}^\prime)$ and $\text{sim}(\mathbf{A},\mathbf{A}^\prime)$ we observe that $K=6$ is the largest number of components that can be extracted while attaining reliable solutions. Inspecting the learned AA representation for $K=6$ we observe that the archetypes provide distinct spectral profiles with their convex combinations defining different regions of the image in which these profiles are present.
 }
\label{fig:AAexampleHyperspectral}
\end{figure}

In remote sensing, AA is motivated by the need to decompose hyperspectral imagery into physically interpretable spectral signatures. For an example of such an application of AA to hyperspectral data, see Figure~\ref{fig:AAexampleHyperspectral}. By identifying extremal spectral profiles and their mixtures across pixels, AA facilitates unmixing, band selection, anomaly detection, and land-cover classification \cite{roscher2015landcover}. Extensions, including kernel AA and ADA \cite{SUN2017147}, further enhance AA's ability to model nonlinear structure and incorporate domain constraints.
\looseness=-1

\footnotetext{Data taken from \url{https://www.ehu.eus/ccwintco/index.php?title=Hyperspectral\_Remote\_Sensing\_Scenes\#Indian_Pines}}

\subsection{Computer and data science}

As a method frequently used in machine learning, AA has been applied to various tasks, not only as a standalone tool but also as an extension to other tools. AA is particularly valuable for tasks requiring explainable representations, such as anomaly detection, clustering, and recommender systems.

\begin{figure}[t]
\centering
\includegraphics[width=\columnwidth]{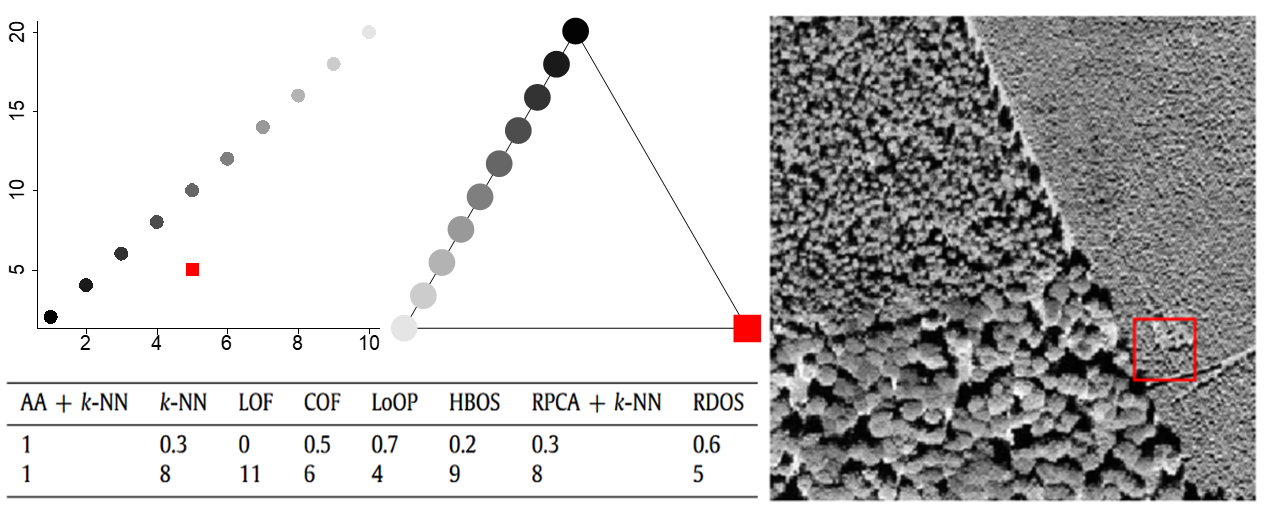}
\caption{On the left, we analyze a two-dimensional dataset where points follow a linear relationship, except for an outlier shown in red. The outlier appears clearly separated from the rest  in the representation for $K$ = 3 AA model, unlike the original space or PCA projection. The table shows AUC results (first row) and the rank of outlierness of the outlier (second row) for the outlier detection methods by \cite{Cabero21}. As there are 11 points, the highest possible rank is 11, which corresponds to the lowest degree of outlierness. Projection in the archetypal space improves distance based outlier detection techniques such as $k$-NN. AA + $k$NN is the only method that detects the outlier. On the right, we reproduce an example by \cite{plos24} where functional AA projections and $k$-NN are used to detect the anomaly in the landscape image.
}
\label{fig:AA_exampleoutliers}
\end{figure}

Its ability to capture extreme but interpretable modes of behavior motivates its use in anomaly detection \cite{MillanEpi}. For examples of the use of AA for outlier detection, see Figure~\ref{fig:AA_exampleoutliers}. From a clustering perspective, AA is applied because it expresses each observation as a convex combination of archetypes, enabling soft, interpretable cluster structures, which is useful in fuzzy clustering \cite{mendes2018study}.
In text analysis, it provides an interpretable alternative to topic models by representing topics as archetypal word profiles \cite{morup2010archetypal}, extending to multi-document summarization \cite{canhasi2014weighted}. In audio and speech, AA is used for representation learning \cite{diment2015archetypal}, while in image and video analysis  it supports feature extraction \cite{Thurau09}  and  concise summarization \cite{song2015tvsum}. In video games, AA is employed to uncover archetypal play styles and behavioral groups for analytics and recommendation \cite{sifa2014archetypal}. Its interpretability also motivates AA-based recommender systems in fashion \cite{math9070771}. Further applications include motion analysis \cite{7346946}, informetrics \cite{wohlrabe2020using}, and software engineering \cite{mittas2014benchmarking}.

Across all these settings, the motivation for applying AA lies in its ability to extract interpretable, extremal patterns, provide transparent analytical insights, and offer flexible, domain-agnostic representations that make complex datasets easier to understand. Besides AA, extensions of AA used in computer science include kernel AA, ADA, weighted AA, functional AA, etc. for capturing nonlinear structure, temporal dynamics, and heterogeneous data common in modern computational tasks.

\begin{figure}[t]
\centering
\includegraphics[width=\columnwidth]{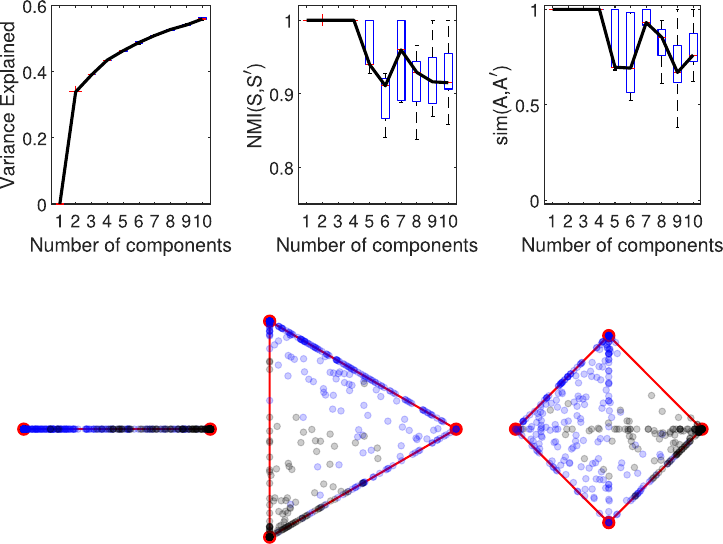}
\caption{Analysis of data of congressional voting records\protect\footnotemark considering 16 bills with votes missing by some congress members on some bills. The Kernel AA procedure is here used considering the Jaccard similarity between congress members in terms of bills they both voted for ignoring bills either had as missing value, i.e., $\mathbf{K}^{\text{Jac.}}_{ij}=\frac{\sum_{m:(x_{im}\in\{0,1\}, x_{jm}\in\{0,1\})}x_{im}x_{jm}}{\sum_{m^\prime:(x_{im^\prime}\in\{0,1\}, x_{jm^\prime}\in\{0,1\})} 1-(1-x_{im^\prime})(1-x_{jm^\prime})}$. The kernel has been double centered, i.e., $\mathbf{TK}^{\text{Jac.}}\mathbf{T}^\transpose$ prior to analyses, corresponding to removing the mean. Black dots correspond to Republicans, whereas blue dots correspond to Democrats. The learned models are illustrated for a $K=2$, $K=3$, and $K=4$ AA model. When inspecting the variance explained and reproducibility as measured using $\text{NMI}(\mathbf{S},\mathbf{S}^\prime)$ and $\text{sim}(\mathbf{A},\mathbf{A}^\prime)$ we observe that up to $K=4$ the AA models can reliably be inferred whereas the variance explained steadily improves as the number of components is increased. Inspecting the model for $K=2$, $K=3$, and $K=4$ we observe that the models well separate Democrats (blue dots) from Republicans (black dots). 
\looseness=-1
}
\label{fig:AAexampleCongress}
\end{figure}
\footnotetext{Data taken from \url{https://archive.ics.uci.edu/dataset/105/congressional+voting+records}}

\subsection{Social sciences and miscellany}
AA has found extensive applications in the social sciences and related fields because it identifies extreme, interpretable patterns (archetypes), that help researchers understand variation, contrast groups, and benchmark performance. In psychology and sociology, AA is used to reveal distinct personality types, identities \cite{fordellone2017finding}, and behavioral profiles \cite{lim2015revealing}, enabling clearer interpretation of complex human traits and social dynamics. In benchmarking and marketing, AA helps identify exemplar performers or representative market segments \cite{li2003archetypal}, supporting decision-making \cite{Porzio2008}. In finance and economics, AA captures extreme corporate profiles or market behaviors, offering insights into structural differences among firms \cite{Moliner2019}. In education and sports analytics, AA highlights distinct student or athlete profiles, aiding evaluation \cite{VinEpi17}, forecasting \cite{VinEpi19}, and personalized interventions \cite{sort20}. Beyond these areas, AA is employed across a range of social science and related fields, where, researchers seek interpretable extremal patterns \cite{WOS001271420100002}. Across these fields, ADA is often preferred because it yields tangible, concrete observations as representatives \cite{wohlrabe2020using}, rather than idealized extremes. As an example of the application of AA to social science data, Figure~\ref{fig:AAexampleCongress} applies kernel AA to congressional voting record data.

\section{Limitations of Archetypal Analysis}\label{Sec:Limitations}

While AA provides interpretable and geometrically meaningful representations, it also has limitations that practitioners should consider. Understanding these constraints is essential for the responsible and effective use of AA.

\textbf{Nonconvex Optimization and Local Minima.}
The standard AA objective in Equation~\eqref{eq:AA_LS_objective} is nonconvex in both variables. Alternating minimization as presented in Algorithm~\ref{alg:AA} may converge to different local minima depending on the initialization. Mitigation strategies include multiple restarts and the use of meaningful initialization schemes (e.g., AA++~\cite{mair2024archetypal}).

\textbf{Sensitivity to Outliers.}
Because AA places the archetypes on the boundary of the data's convex hull, outliers can disproportionately influence the learned archetypes. Robust~\cite{chen:hal-00995911} and weighted~\cite{Eugster2010} variants reduce this effect but at the cost of additional or more involved computations.

\textbf{Choice of the Number of Archetypes.}
Selecting the number of archetypes $K$ remains challenging. Too few archetypes severely underfit the data, while too many archetypes usually reduce interpretability. Model-selection heuristics include scree or elbow plots, explained variance curves, or information-theoretic criteria, but no universally accepted method exists.

\textbf{Scalability and Computational Cost.}
Vanilla AA requires multiple projection steps and large matrix multiplications, resulting in computational costs as shown in Table~\ref{tab:algorithms}. These costs may be prohibitive for large or high-dimensional datasets. Approaches such as reduced-space AA (RSAA) \cite{richardson2021identifying,risbey2021identification,black2022archetypal}, randomized methods \cite{han2022probabilistic}, or coresets \cite{mair2019coresets} alleviate these issues but may introduce approximation errors.

\textbf{Degeneracy and Non-uniqueness.}
If the number of archetypes exceeds the intrinsic dimensionality of the data (e.g., $K > M + 1$), archetypes can become redundant or unstable. Different optimization paths may produce distinct sets of archetypes with similar reconstruction errors, complicating interpretability.

\section{Future directions and open problems}\label{sec:FutureDirandOpenProblems}
In summary, AA has found wide application across diverse scientific domains to characterize datasets based on distinct properties. Although these characterizations have been useful for many different purposes, there are many avenues for future research on AA.

Importantly, as highlighted earlier, the AA model is nonconvex and therefore prone not only to local minima but also to potential unreliability. Whereas this is typically mitigated by using suitable initialization procedures as well as multiple analyses considering different model initializations, future work should investigate the limits of convexity in the AA based on ideas from the geometric properties of convex hulls \cite{kalantari2015characterization} as well as relaxations of AA approaches similar to what has been explored for clustering using spectral relaxations as starting points for the analyses \cite{NIPS2001_d5c18698,von2007tutorial}.

Another open problem is the reliable estimation of the model order, i.e., the number of archetypes $K$. While this is often problem-dependent and requires domain knowledge, having an automated way to determine $K$ would be beneficial. A related open problem is the transfer of knowledge between model orders, i.e., adding or removing an archetype, without recomputing the factorization from scratch.

Some advances in AA stem from the exploration of clustering applications and their extensions \cite{alcacerieee24}. Following this trend, AA could be extended to other types of data and settings, such as censored data, fair AA \cite{alcacer2025incorporating}, or differentially private AA.

Whereas AA has been used to probe temporal dynamics, there are still essential directions for research, including developing methodologies that enable AA to change smoothly over time. Future research directions should thus investigate how ideas from dynamical systems modeling can be combined with AA procedures to characterize the time evolution of polytopes. This includes imposing temporal dependencies on how archetypes ($\mathbf{A}$) may evolve, as well as on how the time-evolving reconstruction of observations $\mathbf{s}_t$ may depend on previous time points, convex combinations, etc. 

An essential limitation of AA is that it relies on the presence of pure forms in the data or that such pure forms can be derived as convex combinations of the observations. In many situations, the AA model cannot recover such pure observations, and the AA modeling assumptions must be relaxed to characterize them. One approach discussed has been to relax the sum-to-one constraints on $\mathbf{c}_k$ by $\delta$. However, this requires suitable tuning of $\delta$, whereas the value of $\delta$ may depend on the specific archetype considered. Future work should investigate how AA procedures can be combined with minimum-volume-based approaches and how to extract polytopes efficiently when the pure forms are not contained within the convex hull of the data. 

Conventional AA assumes that the data resides within a polytope in the considered data space. However, data typically reside on a manifold, and the manifold structure should be taken into account when forming archetypes. Presently, kernel AA can only, to a limited extent, remedy this by imposing a pre-specified kernel structure to account for pairwise relations and the neighborhood structure on the data manifold. Deep AA procedures can learn suitable latent representations in which the data manifold can be disentangled via nonlinear representations. Future work should therefore focus on optimally combining manifold learning approaches with AA.
\looseness=-1

\section{Conclusion}\label{sec:Conclusion}
This survey presented the archetypal analysis (AA) framework, its extensions, and applications across many disparate fields of science highlighting how AA has been successfully applied to characterize and gain insights into the structure of high-dimensional data. 
AA has become a prominent tool for the characterization of distinct aspects in data representing each observation by convex combinations of the extracted archetypes. Consequently, AA excels in providing easily interpretable representations that reveal prominent extremes including their observation specific trade-offs providing a complementary tool to conventional clustering and matrix factorization procedures. We highlighted the many advancements and generalizations of AA and showcased how AA provides interpretable characterizations of data from disparate domains also demonstrating the insights that can be gained from the AA modeling procedure. We at the same time emphasized the importance of quantifying the computational reproducibility by quantifying model robustness to initialization as the inference can be prone to local minima issues. We further discussed the central limitations of the AA procedure as well as important open problems and future directions of research. We hope this will also motivate the community to address some of the current limitations of AA highlighted here, thereby strengthening AA as a data science tool. We further hope this survey will provide an important starting point for researchers unfamiliar with this modeling methodology to adopt this highly explainable procedure as a standard tool in their data analyses, thereby gaining a deeper understanding of the distinct aspects characterizing their data and of how the observations can be described as convex combinations thereof.

\section*{Acknowledgments}

This work was partially supported by the Spanish Ministry of Science and Innovation (PID2022-141699NB-I00 and PID2020-118763GA-I00 to A.A. and I.E.) and Generalitat Valenciana (CIPROM/2023/66 to A.A. and I.E.).

The authors also thank
Adele Cutler,
Ahc\`{e}ne Boubekki,
Anna Emilie Jennow Wedenborg,
Daniel Fern\'andez,
Didier Monselesan,
Samuel G. Fadel, and
the anonymous reviewers
for their valuable comments.

\bibliographystyle{IEEEtran}
\bibliography{references}

\section*{Author bios}

\vspace{-8mm}

\begin{IEEEbiography}
[{\includegraphics[width=1in,height=1.25in,clip,keepaspectratio]{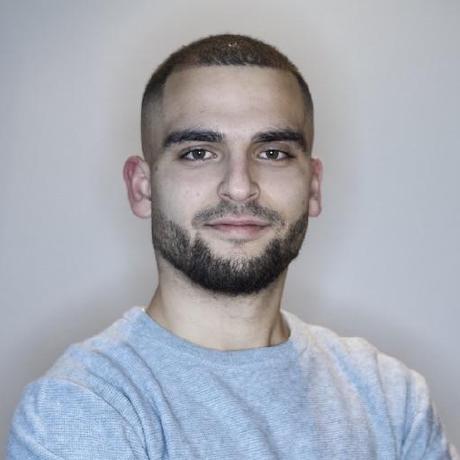}}]
{Aleix Alcacer}
received his BSc and MSc degrees in computational mathematics both from Universitat Jaume I, Spain. He also earned his PhD degree in mathematics from the same institution in 2024. He was awarded the prizes for the best academic records in both his BSc and MSc programs. His research interests include archetypal analysis, data visualization and functional data analysis.
\end{IEEEbiography}

\vspace{-8mm}

\begin{IEEEbiography}[{\includegraphics[width=1in,height=1.25in,clip,keepaspectratio]{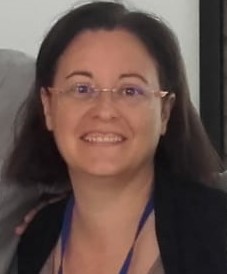}}]{Irene Epifanio} received her MS degree in mathematics and her PhD degree in statistics from València University, Spain. She is currently a full professor of Statistics at Jaume I University, Spain, and Senior Research Fellow at valgrAI. Her research interests include statistical learning, functional data analysis, computer vision and equality. She was the recipient of various awards in research, teaching, scientific dissemination and social commitment, including the Margarita Salas Prize of Talent Woman Spain.
\end{IEEEbiography}

\vspace{-8mm}

\begin{IEEEbiography}[{\includegraphics[width=1in,height=1.25in,clip,keepaspectratio]{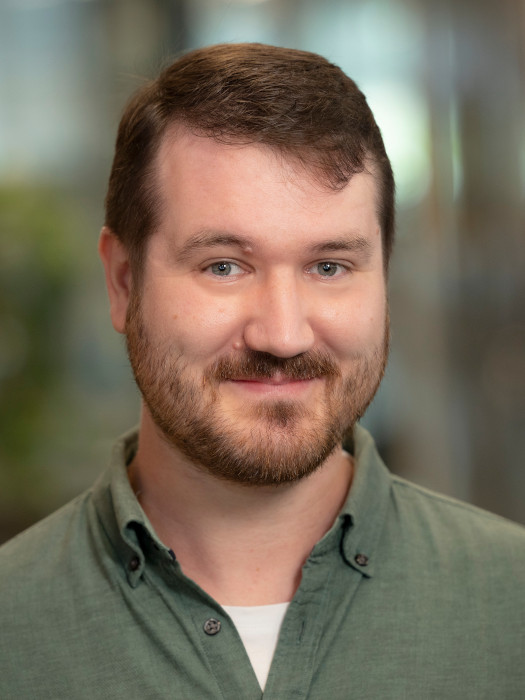}}]{Sebastian Mair}
received his BSc degree in mathematics and MSc degree in computer science both from Technical University of Darmstadt, Germany and his PhD degree in computer science from Leuphana University of Lüneburg, Germany. After his PhD, he was a postdoctoral researcher at Uppsala University, Sweden. Currently, he is an assistant professor of statistical machine learning at Linköping University, Sweden. His research interests lie in the span of unsupervised learning, representation learning, representative subsets, generative modeling, and trustworthy and sustainable machine learning.
\end{IEEEbiography}

\vspace{-8mm}

\begin{IEEEbiography}
[{\includegraphics[width=1in,keepaspectratio]{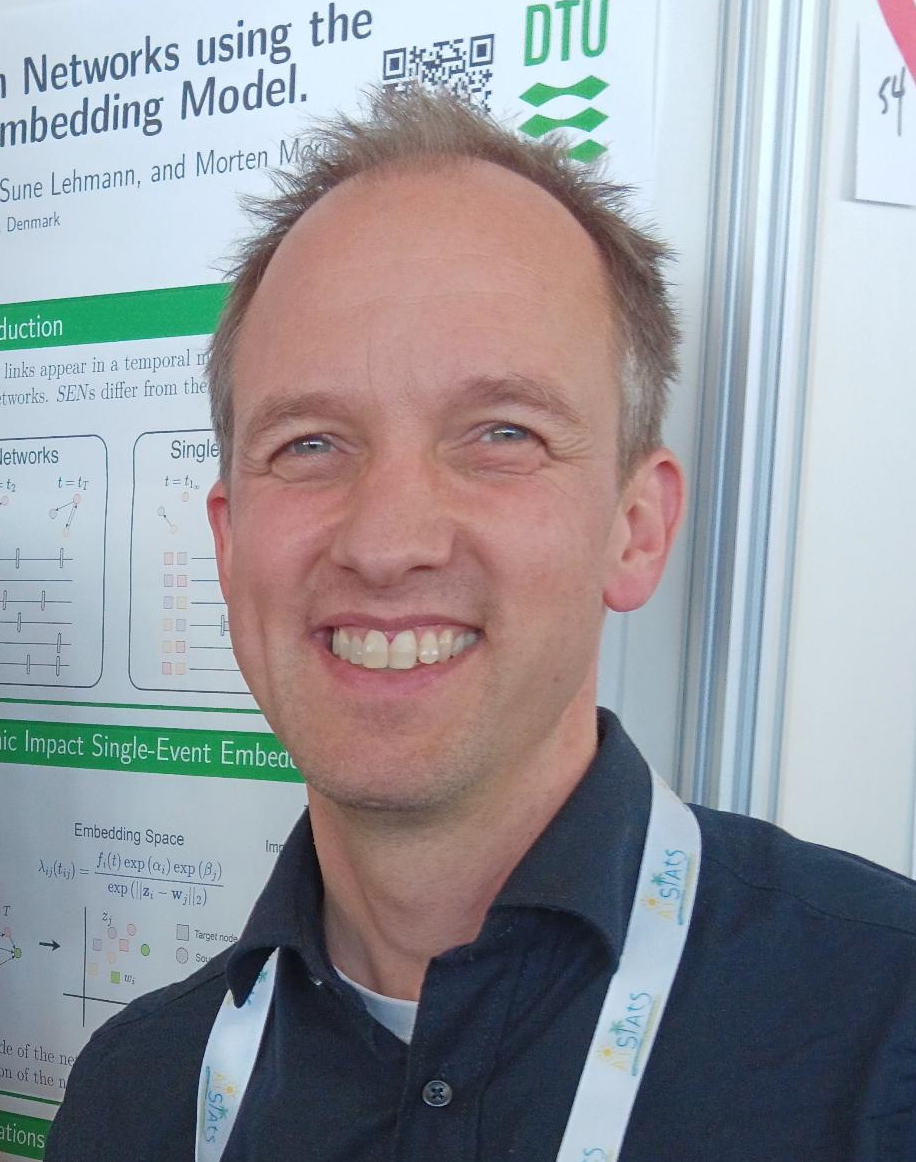}}]{Morten M{\o}rup}
 received his MS and PhD degrees in applied mathematics from the Technical University of Denmark, Denmark, where he is currently a professor of machine learning for the life-sciences at the Section for Cognitive Systems at DTU Compute. He has been associate editor of the IEEE Transactions on Signal Processing, area chair for multiple leading machine learning conferences, and his research focuses on machine learning, tensor decomposition, and complex network modeling for the modeling of life-science data in general and neuroimaging data in particular.
\end{IEEEbiography}

\vfill

\newpage

\appendices

\section{Prominent applications of AA}\label{sec:ProminentApplicationsofAA}

This appendix presents an extended and more detailed version of Section~VII.

\subsection{Life science}
AA has found wide applications in the life sciences. As such, the extraction of polytopes has been applied to biological data to characterize evolutionary trade-offs \cite{shoval2012evolutionary} in which archetypes define optimal traits for various tasks. 

Inspired by such extraction of traits, AA has been used within the domain of genetics \cite{Morup2013,hart2015inferring, szekely2015mass,milite2025midaa,groves2022archetype,adler2023emergence,korem2015geometry}, including for the characterization of prominent genetic profiles at a population level \cite{gimbernat2022archetypal}. AA has further been applied to single-cell Ribonucleic acid sequencing (scRNA-seq) data to characterize cell diversity \cite{mohammadi2018geometric}, also considering kernel AA based on RNA-seq cell similarities \cite{persad2023seacells}, as well as nonlinear representations considering a deep AA approach \cite{wang2022non}. In proteomics, AA was used to identify extreme protein-expression patterns, guiding active machine-learning experiments to determine compound effects on cellular protein profiles \cite{naik2016active}. In \cite{reeve2017assessing}, microarray data from kidney transplant biopsies were scored in terms of aspects of organ rejection, and AA was used to characterize the molecular phenotypes related to kidney transplant rejection. In \cite{demir2024identification} liver transplant biopsies with donor-specific antibody testing were analyzed by AA for the characterization of phenotypes associated with distinct liver properties and transplant survival. Similarly, \cite{halloran2018exploring} used AA to identify extreme cellular and molecular response patterns in heart transplant biopsies, revealing distinct profiles of cardiac injury and repair.

In neuroimaging, AA has been used to characterize low-binding, high-binding, and non-binding regions of tracers in positron emission tomography (PET) studies \cite{morup2012archetypal}. Furthermore, AA has been used to characterize distinct functional activation patterns in functional magnetic resonance imaging (fMRI) data \cite{hinrich2016archetypal,krohne2019classification}, including responses to visual stimuli \cite{hinrich2016archetypal,ravindra2021characterizing} and for the characterization of single-trial variability in magneto-encephalography (MEG) data \cite{tsanousa2015novel}. It has further been advanced to accommodate multimodal integration of electro- and magneto-encephalography (EEG and MEG) stimuli-evoked responses \cite{olsen2022combining,olsen2024coupled}, providing distinct topographic patterns as an alternative to conventional micro-state analysis approaches \cite{olsen2022combining}. In \cite{beck2022archetypal}, AA was applied to calcium fluorescence imaging for the detection of groups of neurons systematically co-activating.
\looseness=-1

AA has been further applied in the context of medicine and health. As such, AA was used to characterize the spatio-temporal dynamics of influenza outbreaks \cite{mokhtari2021decoding} and COVID-19 \cite{stone2024archetypal} in Montana, USA. In \cite{grane2022looking} ADA was applied to characterize effects on health of COVID-19 across Europe, whereas European countries' epidemiological data on COVID-19 at two time points were analyzed by AA in \cite{vicente2022covid}. In \cite{trescato2024dynamite}, the disease progression of amyotrophic lateral sclerosis patients was examined by their dynamic characterization in terms of archetypes describing various aspects of impairments. AA has also been used to help characterize extreme biomechanical response patterns of the optic nerve head under varying cerebrospinal fluid pressure, offering interpretable summaries of ocular tissue behavior \cite{hua2018cerebrospinal}. 
Notably, visual field (VF) loss has been extensively characterized using AA \cite{doshi2021unsupervised,doshi2022unsupervised,szanto2024archetypal} and related to established clinical features \cite{elze2015patterns}, and specifically for the characterization of optic neuritis in \cite{thakur2020convex} and glaucoma in \cite{cai2017clinical,mahotra2020patterns,singh2024machine}. In this context, a deep learning-based AA approach was considered in \cite{solli2022archetypal}, and a hierarchical recursive approach in which the convex combinations extracted for observation reconstruction were recursively decomposed by AA was considered in \cite{mahotra2020patterns,gupta2020glaucoma}. More recently, \cite{coutinho2025machine} applied machine-learning–derived archetypal patterns to visual fields of patients with dominant optic atrophy, further broadening the clinical scope of AA-based VF characterization. 

Other biomedical applications of AA include the following. \cite{ortigueira1999archetypal} used AA to examine and interpret ECG signals. In audiology, \cite{sanchez2018data} identified archetypal listening profiles, allowing individualized characterization of complex hearing-loss patterns. In traumatology, \cite{aubert2024archetype} applied AA to the spine-hip relationship to reveal distinct spinopelvic profiles. In cancer immunology, \cite{hes2023gut} identified extreme gut microbiome and nutrition-related profiles, helping to interpret and predict individual responses to immunotherapy. In child development research, \cite{han2018functional} found archetypes of growth and identified subgroups related to childhood cognitive development. \cite{Yelugam2025ExplanationSpace} used BiAA for clustering of neurologic diseases based on their signs and symptoms. In \cite{pancotti2023unravelling} AA was used to characterize redundancies within a catalog of somatic mutations in cancer and in \cite{trescato2023identifying} to identify extreme clinical types of amyotrophic lateral sclerosis (ALS) patients. Furthermore, AA has been used to characterize reward strategies in mice \cite{10.1371/journal.pbio.3002850}. AA has also been used to analyze survey data on the physical activity of pregnant women \cite{karwanski2024archetype}. 

Finally, AA has also been applied in the study of ergonomics to characterize foot shapes \cite{EpiVinAle}, anthropometric measurements relevant for the design of aircraft cockpits \cite{plos20} and exoskeletons \cite{rodriguez2021comparison} as well as to identify boundary cases for ergonomic workstation designs defined in terms of the nearest neighbors to the identified archetypes \cite{riemer2022study}.

\subsection{Physics and chemistry}
AA has been applied to different problems in physics and chemistry. They range from the analysis of the largest entities in the universe, studied in astrophysics, to the smallest, studied in nanotechnology. \cite{chan2003archetypal} employed AA to examine datasets consisting of galaxy spectra, as each spectrum is thought to be a superposition of emissions from various stellar populations, nebular emissions, and nuclear activity within the galaxy. Each of these emission sources represents a potential archetype for the entire dataset. Moreover, data from the Tokamak Fusion Test Reactor was analyzed with AA by \cite{cutler1994archetypal}. AA has also been used to analyze data from dynamical systems \cite{stone2002exploring,cutler1997moving,stone1996archetypal,stone1996introduction}, in a chemical pulse experiment \cite{cutler1997moving}, a numerical simulation of the Kuramoto–Sivashinsky equation \cite{stone1996archetypal}, and cellular flame data \cite{stone1996introduction,stone2002exploring}. Furthermore, AA has been applied by \cite{subramanian1998sampling} in wafer misregistration.

We also find AA applications in chemistry, such as the analysis of 
$^1$H NMR spectra of mixtures of propanol, butanol, and pentanol by \cite{morup2012archetypal} and mid-infrared spectral library of soil samples by \cite{sila2016evaluating}. Moreover, virtual samples of diamond nanoparticles and graphene nanoflakes are analyzed using AA by \cite{fernandez2015identification} and graphene oxide nanoflakes by \cite{motevalli2019representative}, while archetypal nanostructures of five virtual ensembles of Si quantum dots (SiQDs) are described by \cite{fernandez2017impact}. Finally, AA has been applied for the chemical analysis of wines \cite{santelli2019statistical} and interpretation of elemental distribution images obtained via X-ray fluorescence acquired on historical paintings \cite{alfeld2014non,alfeld2017simplex}.

\subsection{Climate science and sustainability}
AA has been utilized in applications related to climate science. For instance, \cite{steinschneider2015daily} studied daily precipitation and tropical moisture exports across the eastern United States; the Asian summer monsoon using sea level pressure is analyzed by \cite{hannachi2017archetypal} together with the monthly sea surface temperature, as done by \cite{black2022archetypal}, \cite{monselesan2024archetypalflavours} and \cite{chapman2022large}; long-lived Southern Hemisphere flow events are examined by \cite{risbey2021identification}; large-scale atmospheric circulation regimes that lead to summertime heatwaves in the Northern Hemisphere are studied in \cite{chapman2025typicality}; periods of high confidence in subseasonal forecasts were identified by \cite{richardson2021identifying}; \cite{su2017spatiotemporal} investigated the spatio-temporal variability of seasonal extreme precipitation; \cite{Thurau12} studied historical climatography series temperature and precipitation normals; finally, \cite{monselesan2024extreme} studied archetypes of extreme rainfall in Australia and published the related datasets.
Data mining approaches, including archetypal analysis for weather and climate science, were also summarized and explained in \cite{hannachi2021patterns}.

Besides climate data, AA has been used for other computational sustainability problems. As such, \cite{Thurau12} analyzed global electricity consumptions whereas air pollution was studied in the seminal problem that originated AA \cite{cutler1994archetypal}; \cite{novotny2023looking} analyzed households' food quality and sourcing in Togo; \cite{taeger2013impact} explored the growth response of Scots pine to climate and drought events; \cite{wang2022sugarcane} predicted sugarcane biomass; \cite{easdale2024tracking} studied floristic archetypes of Patagonian steppes; \cite{tessier2021identifying} analyzed the farming models that underpin the practices of Flemish beef farmers through an agroecological lens, while \cite{juventia2025sole} identified extreme cropping strategies among frontrunner farmers in the Netherlands; \cite{vidan2019global} examined the global biogeography of functional groups within lizard populations; \cite{reyes2017non} studied volcanic-seismic events for Tungurahua-Volcano Ecuador; \cite{pecuchet2017traits} explored fish community composition across European seas; \cite{theodosiou2012measuring} analyzed \textit{Olea europaea} production data; \cite{abbasiharofteh2024using} analyzed the relational properties within European bioclusters' knowledge networks; \cite{ferdinand2024method} applied AA to capture and summarize the diversity of conservation agriculture practices; in \cite{tocchi2025identifying}, AA was applied to classify urban and rural settlements into representative archetypes, supporting more targeted risk-oriented exposure and vulnerability assessments; \cite{hitt2022life} characterized stream-fish life histories as mixtures of extreme strategy types linked to hydrologic stability; \cite{lapierre2023continuous} extracted lake archetypes to better depict lake states and anticipate how they will respond to environmental change; \cite{bains2025norepinephrine} identified differences in metabolic pathways between dextrose and L-NE growth conditions in wastewater microbiomes; \cite{robinson2025hydrological} found hydrological archetypes of wetlands; \cite{arbelaezarchetype} analyzed golden eagle migration patterns; finally, \cite{MillanEpi} worked on the detection of anomalous flows in urban water networks.

AA has also been used in the context of remote sensing for the analysis of hyperspectral satellite imagery. In \cite{Thurau12,romer2012early}, hyperspectral images were analyzed using a simplex-volume minimization procedure and AA finding that these procedures provide easily interpretable representations of the underlying spectral profiles. In \cite{roscher2015landcover}, this approach was found to provide features well suited for land cover classification, whereas the approach extended to kernel AA in \cite{zhao2016multilayer,zhao2016hyperspectral} using a radial basis function kernel and to archetypes representing actual cases (i.e., archetypoids) in \cite{SUN2017147}. In \cite{sun2017probabilistic}, the Poisson likelihood was used for AA to account for integer-valued pixel intensities imposing a weighted framework based on earth mover distances (EMD) between the observations when forming the archetypes whereas AA was used in \cite{sun2016symmetric,sun2018band} for the selection of hyperspectral bands suitable for classification and comparison of plant-disease dynamics by \cite{wahabzada2015metro}. \cite{ni2017fast} allowed the archetypes to be any representative samples. In \cite{drees2018archetypal}, a reversible jump Markov Chain Monte Carlo inference procedure combined with simulated annealing was used to identify globally robust representations of archetypes for hyperspectral images based on candidate AA representations. Furthermore, in \cite{zouaoui_entropic_2023}, the entropic gradient descent was explored to provide efficient explicit updates enforcing the mixture constraints in AA for the analyses of hyperspectral images. \cite{wang2025inferring} proposed a simplex-projected gradient descent formulation of archetypal analysis. In \cite{rasti2023sunaa} the archetypes were defined in terms of convex combination of endmembers obtained from a library of predefined spectra and in \cite{xu2023manifold} the convex constraints on $\mathbf{C}$ relaxed using the relaxation approach proposed in \cite{morup2012archetypal} in combination with a regularization imposing similar reconstruction of pixels belonging to the same superpixel segment. In \cite{bruzzone2024tracking}, the Normalized Difference Vegetation Index (NDVI), defined by the normalized difference between the near-infrared and red bands, was analyzed by transforming the spatio-temporal data to power spectra that served as input to AA based on a continuous wavelet transform. In \cite{zhao2021archetypal}, AA was used for anomaly detection in hyperspectral images.

\subsection{Computer and data science}

As a method frequently used in data science and machine learning, AA has been applied to various tasks, not only as a standalone tool but also as an extension to other tools.
\looseness=-1

One of the core properties of AA is its ability to identify \emph{extreme} prototypes that live on the boundary of the dataset.
This makes AA a perfect candidate for anomaly and outlier detection.
For example, AA was used for offline handwritten signature verification \cite{zois2017offline} and to support DevOps in their work.
\cite{9140432} constructed an AA-based one-class classifier for cyber-physical systems, whereas \cite{MillanEpi} used functional data analysis with AA and ADA for anomaly detection in water networks.
ADA was also used by \cite{Vinue21}, e.g., for time series data such as ECG data.
\cite{Cabero21} proposed to combine AA and a nearest neighbor algorithm to obtain a new subspace outlier detection approach and \cite{plos24} extended their approach. 
Finally, \cite{sifa2021archetypal} used AA for anomaly detection in electronic game analytics.

As outlined earlier, AA can also be seen from a clustering perspective, for which \cite{mendes2018study,nascimento2019unsupervised,suleman2021comparing} evaluated AA for fuzzy clustering.
Moreover, \cite{tan2022reusable} used AA as a clustering approach to group video game players based on their player behavior, using representations obtained from a recurrent autoencoder.

AA is frequently used to analyze data from different modalities.
For the analysis of text data, \cite{morup2010archetypal,morup2012archetypal} used AA as a topic model in the spirit of latent Dirichlet allocation (LDA) \cite{blei2003latent}.
The idea is to extract a set of topics from a collection of texts, where each topic is represented by archetypal words.
\cite{steuber2020topic} proposed an AA-based extension of LDA.
Another line of research in this area deals with multi-document summarization, where various variants of AA are used \cite{canhasi2014multi,canhasi2014weighted,canhasi2015automatic,canhasi2016weighted}.

For audio data, \cite{diment2015archetypal,virtanen2018separation} used AA as a tool for representation learning and for source estimation and \cite{sinni2022singing} for singing voice separation. 
For audio classification, \cite{singh2018ape} combined archetypes as boundary representations and centers from Gaussian mixture models to obtain better embeddings for a random forest classifier, and \cite{thakur2019deep} proposed an AA-based intermediate matching kernel to train a support vector machine (SVM) for the classification of bioacoustic data. 

Another modality is image data.
\cite{Thurau09} extracted features of images and compute an AA on those features to obtain archetypal images.
\cite{morup2010archetypal,morup2012archetypal} ran an AA directly on image data showing human faces.
\cite{chen:hal-00995911} performed AA-based classification of images showing digits.
For another task, \cite{chen:hal-00995911} split images into patches, computed SIFT features per patch, and computed visual words using AA to provide a representation for classifiers.
The idea of splitting images into patches was also used by \cite{bauckhage2015archetypal}.
Instead of computing features per patch, they used the patches as they are to compute archetypes of patches.
Then, patches of images were approximated by convex combinations of archetypal patches, yielding a kind of autoencoder. AA is also used for identifying a small set of ``virtual markers'' that summarize the full distribution of human body shapes and poses \cite{ma2025vmarker}.
Finally, \cite{Xiong_2013_ICCV} computed archetypes for face recognition tasks, not exactly via AA, but via a related approach called simplex volume maximization \cite{thurau2010yes}. This is also used by \cite{kersting2012simplex} to embed data matrices over time. More image applications of AA are found in \cite{potts2021adapting}.

As a last modality, we consider video data for which \cite{song2015tvsum} used AA to summarize videos based on a set of features, as well as the textual information of the title, and \cite{ji2019multi} performed multi-video summarization via multimodal weighted AA.

Another prominent use case of AA is video games.
There, \cite{sifa2013archetypical} extracted movement primitives using AA to be used for game bot creation, \cite{sifa2014archetypal} proposed an AA-based recommender system for video games, \cite{sifa2014playtime} evaluated the playtime behavior by kernel AA, \cite{schiller2018inside} leveraged AA to understand groups of players in video games, \cite{sifa2021archetypal} performed video game analytics, \cite{tan2022reusable} used AA as a clustering approach to group video game players by their player behavior based on representations obtained from a recurrent autoencoder, and \cite{su2024better} employed AA to find archetypes of play styles in games that are used to train an agent per archetype that optimally cooperates with a human. \cite{chitayat2025ai} analyzed data on Twitch’s most successful games to identify distinct success archetypes.

AA is also used to build recommender systems.
\cite{morup2010archetypal,morup2012archetypal} used AA for collaborative filtering on the movielens data.
There, the task is to infer missing values in user ratings for movies.
An AA-based recommender system for video games was proposed by \cite{sifa2013archetypical} and \cite{math9070771} proposed a recommendation system combining classification and user-based collaborative filtering using AA for the fashion industry. 

AA has been further considered within the fields of motion analysis and map design.
\cite{sifa2013archetypical} used AA to extract movement primitives in video games to be used for game bot creation.
A temporal AA for the segmentation of actions in multimodal times series describing human motion is proposed by \cite{fotiadou2017temporal}.
\cite{7346946} and \cite{feld2017identifying} used AA to evaluate indoor routes and \cite{bartling2021modeling} used it to evaluate map design usability. AA is also used in cognitive radio \cite{balaji2016cooperative}.

In recent explainable AI research \cite{wedenborg2026explaining}, 
AA is used to interpret the latent (internal) feature spaces of neural networks with the aim of describing how information is represented layer-by-layer inside the model.

As regards benchmark creation for scientific datasets, \cite{barnard2025benchmake} detected challenging edge cases using AA. Another use of AA is as a sample augmentation method and as an imputation method \cite{cavalcanti2021archetypal}.

Within informetrics, \cite{ramos2022value} used AA to understand the interplay between science and society and especially the value creation aspect of it.
\cite{seiler2013archetypal} extracted six archetypes of economists based, e.g., on bibliographic indicators with the help of AA.
This study was extended by \cite{gralka2022classifying} to ADA.
\cite{wohlrabe2020using} also leveraged ADA to extract three archetypal institutions and faculties within the field of economics.
Last, \cite{bauckhage2014kernel} used kernel AA to cluster the dynamics of collective attention in social media.

Finally, AA was used in software engineering, e.g., to identify prototypical models for the task of software effort estimation \cite{mittas2014benchmarking,mittas2020data}, to analyze technical debt \cite{amanatidis2020evaluating}, and for the analysis of contribution types of open source software developers \cite{ikemoto2020analysis}.

\subsection{Social sciences and miscellany}
AA has found extensive applications in the social sciences. In the field of psychology, \cite{stoker2023configurational} examined archetypes of leadership, \cite{kosti2016archetypal} investigated archetypal personalities and \cite{fordellone2017finding} delineated personality traits, while \cite{lim2015understanding,lim2015revealing} analyzed social identities and behavioral archetypes, as well as \cite{drachen2012guns,sifa2013behavior, bauckhage2014clustering,sifa2018overview} who studied behavioral profiles in digital games or \cite{swait2018modeling} in consumer behavior. Furthermore, \cite{ragozini2017archetypal} used AA to study identity formation in adolescence. \cite{ragozini2017archetypal} also considered AA in another field of application, such as benchmarking, where benchmark performers among operative units of a telecommunications company were found. \cite{Porzio2008} also proposed the use of AA to identify benchmark universities and \cite{d2008new} to rank multivariate
performances. Alike, \cite{antoniadis2023benchmarking} applied AA to discover benchmarking classifiers for loan default prediction. In the related field of marketing, AA has also been used to analyze operations of multinational corporations by \cite{midgley2013marketing,venaik2019archetypes} and to segment markets by \cite{li2003archetypal}. In business process management, \cite{fehrer2025taxonomy} derived and evaluate archetypes of business process improvement and innovation systems. Moreover, AA has been applied in finance and economics. For example, \cite{moliner2018bivariate,Moliner2019} described companies in the S\&P 500 depicted by two time series of stock quotes and \cite{grzybowska2022archetypal} enterprises by several financial indicators, while \cite{stehlikova2024clustering} analyzed the zero-coupon yield curve in the USA and \cite{hayden2018canonical} identified canonical sectors in the USA market. Furthermore, in economic geography, \cite{abbasiharofteh2025average} applied archetypal analysis to uncover extreme spatial-economic profiles. 

There are also applications in Sociology. For instance, \cite{beugelsdijk2022nature} examined identities in European countries to study the societal conflict. Another example is the analysis of time use of academics carried out by \cite{cabero2023analysis}. 

Furthermore, AA has been applied within the education field. For instance, \cite{theodosiou2013courseware} analyzed course usage and log files of online courses and their assessment \cite{kazanidis2016online}, while student workload at home and student responses to teacher surveys were analyzed by \cite{epifanioCARGAS} and \cite{epifanioanalisis}, respectively. Finally, \cite{sort20} analyzed binary questionnaires to identify student skill set profiles and explain the items of a math exam. \cite{adabbo2021statistics} also grouped students by archetypes.

Sport analytics is another field of AA application. On the one hand, \cite{Eugster2012} applied AA to the analysis of several variables of basketball and soccer players. On the other hand, \cite{VinEpi17} used ADA, not only for the analysis of several variables of basketball players, but also for studying player career trajectories with functions and team performance through the information of asymmetric relations. Furthermore, ADA for sparse functions was used to forecast basketball players’ performance by \cite{VinEpi19}.

There are also applications in other fields, such as speech evaluation \cite{gu2020characterization} and modeling dialectal variation \cite{canhasi2023using}, analyzing cultural models and their implications \cite{midgley2018culture,harrell2017culturally,venaik2015mindscapes}, cultural transmission \cite{macdonald2024cultural}, artistic style analysis \cite{wynen2018unsupervised}, analysis of terrorist events \cite{Lundberg02092019}, thermal nondestructive testing/evaluation \cite{marinetti2006matrix,marinetti2007archetypes}, social media analytics \cite{charmanas2025exploring}, Toronto housing \cite{ergun2014potential} and architectural design \cite{Chen2015AnalysingPO}. 

Moreover, AA and their variants have been applied for analyzing very diverse survey data, such as interval-coded sensory data \cite{d2011use}, financial knowledge data \cite{palazzo2024integrated}, community attachment data \cite{cutler3}, gender equality actions in university data \cite{WOS001271420100002}, European Social Survey data \cite{wedenborg2024modeling} and quality of life data of patients affected by breast cancer, and student satisfaction survey data \cite{FERNANDEZ2021281}.

\end{document}